\documentclass[twocolumn]{aastex61}
\usepackage{natbib}

\begin{document}

\title{WIRED for EC: New White Dwarfs with \emph{WISE} Infrared Excesses and New Classification Schemes from the Edinburgh-Cape Blue Object Survey}

\author{E. Dennihy}
\affil{Physics and Astronomy Department, University of North Carolina at Chapel Hill, Chapel Hill, NC 27599}

\author{J. C. Clemens}
\affil{Physics and Astronomy Department, University of North Carolina at Chapel Hill, Chapel Hill, NC 27599}

\author{John H. Debes}
\affil{Space Telescope Science Institute, Baltimore, MD 21218}

\author{B. H. Dunlap}
\affil{Physics and Astronomy Department, University of North Carolina at Chapel Hill, Chapel Hill, NC 27599}

\author{D. Kilkenny}
\affil{Department of Physics, University of the Western Cape, Private Bag X17, Bellville 7535, South Africa}

\author{P. C. O'Brien}
\affil{Physics and Astronomy Department, University of North Carolina at Chapel Hill, Chapel Hill, NC 27599}

\author{J. T. Fuchs}
\affil{Physics and Astronomy Department, University of North Carolina at Chapel Hill, Chapel Hill, NC 27599}

\begin{abstract}
We present a simple method for identifying candidate white dwarf systems with dusty exoplanetary debris based on a single temperature blackbody model fit to the infrared excess. We apply this technique to a sample of Southern Hemisphere white dwarfs from the recently completed Edinburgh-Cape Blue Object Survey and identify four new promising dusty debris disk candidates. We demonstrate the efficacy of our selection method by recovering three of the four \emph{Spitzer} confirmed dusty debris disk systems in our sample. Further investigation using archival high resolution imaging shows \emph{Spitzer} data of the un-recovered fourth object is likely contaminated by a line-of-sight object that either led to a mis-classification as a dusty disk in the literature or is confounding our method. Finally, in our diagnostic plot we show that dusty white dwarfs which also host gaseous debris lie along a boundary of our dusty debris disk region, providing clues to the origin and evolution of these especially interesting systems. 
\end{abstract}


\section{Introduction}

The now firmly established link between remnant exoplanetary systems, compact circumstellar dust disks, and atmospheric heavy metal pollution in white dwarf stars has demonstrated that exoplanetary science has much to gain by continuing studies post-mortem. In contrast with their dead host stars, the surviving exoplanetary systems are dynamically active, with large outer planets scattering smaller rocky bodies into disruptively compact orbits, providing a source of rocky material for both the observed compact debris disks and the otherwise unexpected atmospheric metals \citep{deb02,jur08,fre14,ver16}. These post main-sequence exoplanetary systems have revealed detailed rocky exoplanetary abundances \citep{zuc07,duf12,gan12,xu14}, complex debris disk dynamics \citep{wil14,man16a}, and even transits of actively disrupting bodies \citep{van15,rap16}. Despite these exciting discoveries, statistical analyses of the white dwarfs with observable exoplanetary debris disks suggest our picture is far from complete. The frequency of white dwarfs with exoplanetary systems as observed via accretion of heavy metals in white dwarf atmospheres is highly discrepant from the frequency of observable debris disks around white dwarf stars (see \cite{far16} for a recent review), suggesting that majority of the sources of the accreting material remain undetected \citep{roc15,bon17}. The continued discovery of new white dwarf debris disk systems will expand the range of exoplanetary accretion phenomena we observe, and once our understanding of their origin and evolution is mature, allow us to translate the observed sample properties through the initial-final mass relationship to improve our understanding of planetary formation around stars like our sun (e.g. \cite{bar16}). The large number of new white dwarf stars expected to be discovered by the \emph{Gaia} space telescope offers the chance to increase the sample of white dwarf exoplanetary systems by more than an order of magnitude, once the dusty debris disk systems can be identified and confirmed \citep{rob12,car14,gan16}.

The process of identifying candidate dusty debris disk systems in the modern astronomical era of cross matching and database mining is simple: given a list of white dwarf stars, a spectral energy distribution can be constructed entirely from publicly available photometry, then compared against atmospheric models to reveal targets with excess infrared radiation. Unfortunately the infrared signature of dusty debris disks is only prominent beyond 2$\,\mu$m, forcing data miners to rely heavily on the all-sky survey of the Wide-field Infrared Survey Explorer (hereafter \emph{WISE}, \cite{wri10}). While the photometry from \emph{WISE} is well calibrated, its large imaging beam leads to a high probability of un-resolved source contamination \citep{deb11,bar14}, and the two longest wavelength survey passbands, \emph{W3} and \emph{W4} at 12 and 22$\,\mu$m, are not sufficiently deep for studies of white dwarf debris disks, leaving us with only one or two significant points to discern the excess. In the past astronomers have turned to space-based observatories such as the \emph{Spitzer} Space Telescope to confirm white dwarf debris disk candidates at higher spatial resolution and longer wavelengths than can be observed from the ground, which has continued to provide observations well into its warm mission \citep{bar12,xu12,ber14,roc15}. This has worked well for small samples of interesting targets, but is impractical for the hundreds of new dusty infrared excess candidates expected from \emph{Gaia}. 

In this work, as an extension of the \emph{WISE} InfraRed Excess around Degenerates Survey (hereafter WIRED, \cite{deb11}), we present a simple but robust method for identifying the most promising dusty debris disk candidates based on the best-fitting effective temperature and radius of a single temperature blackbody. This technique is well suited to handle the sparse infrared excess points for studies that rely on \emph{WISE} photometry. We apply this technique to the set of hydrogen atmosphere (DA) white dwarfs identified in the recently completed Edinburgh-Cape Blue Object Survey (hereafter EC Survey, \cite{sto97}, \cite{kil16}). This sample of Southern Hemisphere white dwarfs is relatively bright (\emph{V} $\leq$ 17.5) and provides a good proxy for some of the issues that can be expected and will need to be overcome for the sample of bright white dwarfs from \emph{Gaia}, which will initially have little spectroscopic data and varying photometric coverage. 

We present four new, promising dusty debris disk candidates, and identify a known dusty debris disk hosting white dwarf as an outlier with our technique, which we show to have a nearby contaminant that is unresolved in previous \emph{Spitzer} studies. We also find that among the \emph{Spitzer}-confirmed dusty debris disk hosting white dwarfs, those that also host an observable gaseous component lie along the boundary of our dusty debris disk selection region, confirming a relationship between the infrared disk luminosity and its propensity to host gas. 

\section{Target Selection and Collected Photometry}

\begin{figure*}[t]
\gridline{\fig{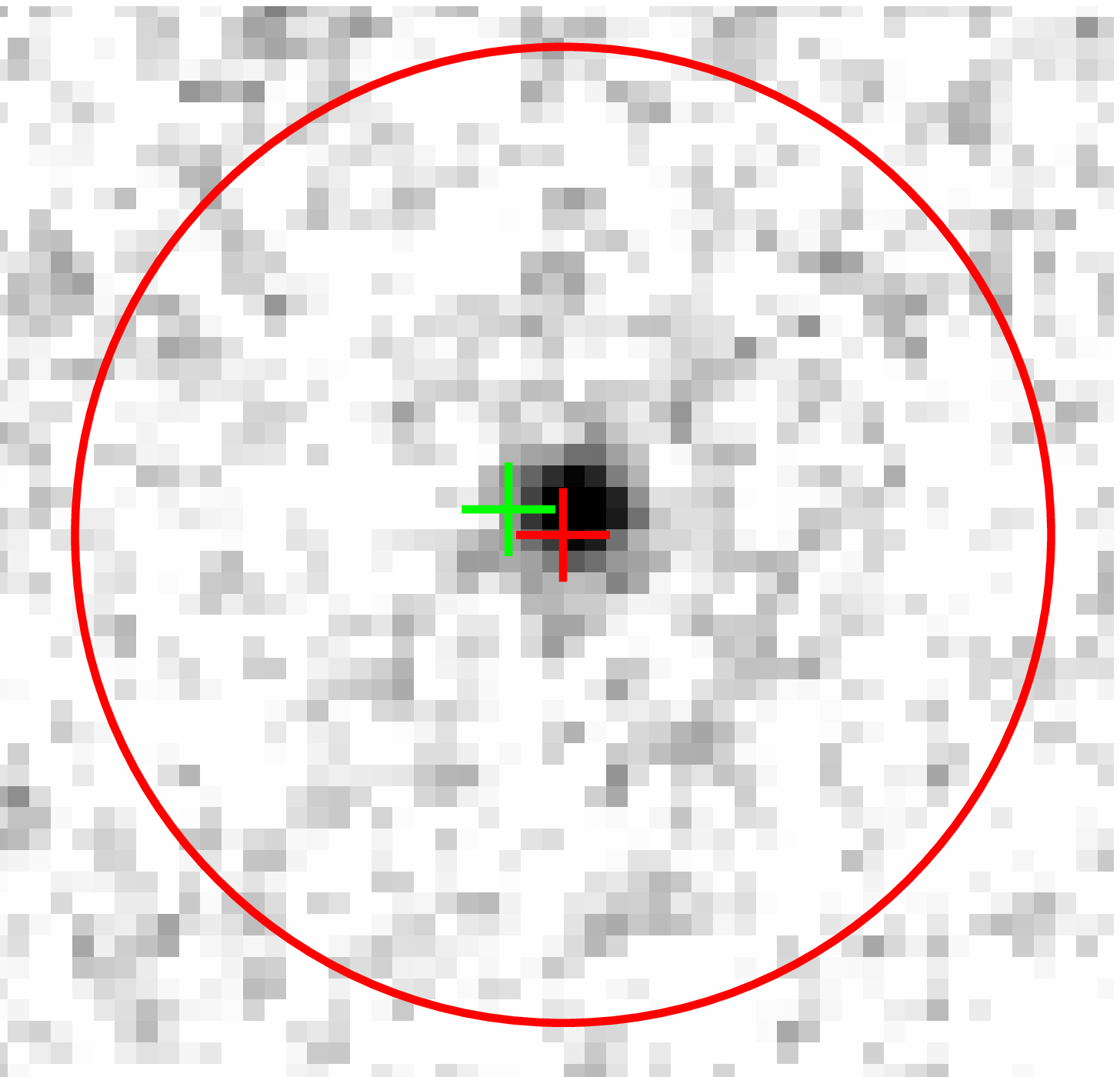}{0.25\textwidth}{(a)}
          \fig{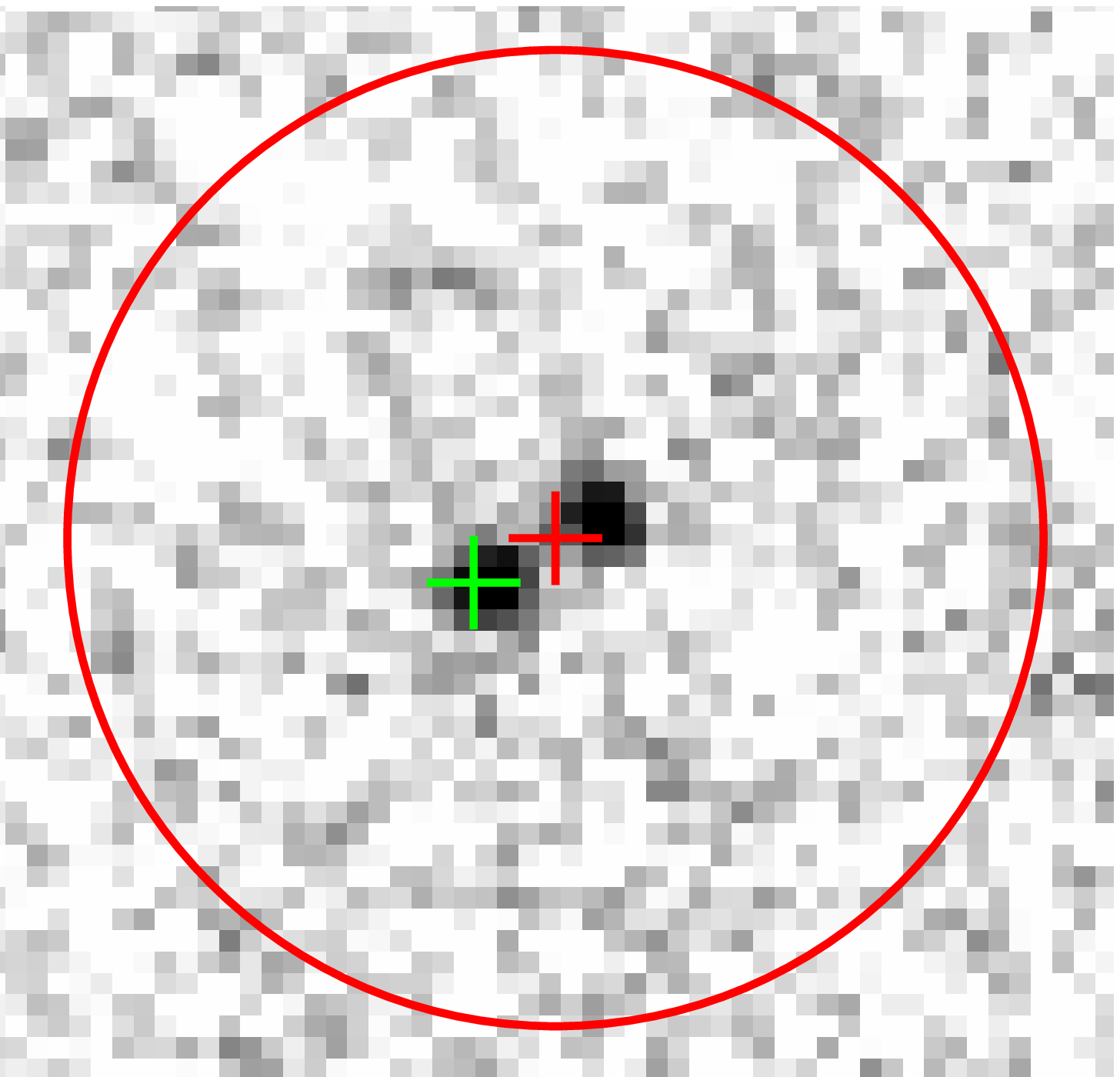}{0.25\textwidth}{(b)}
          \fig{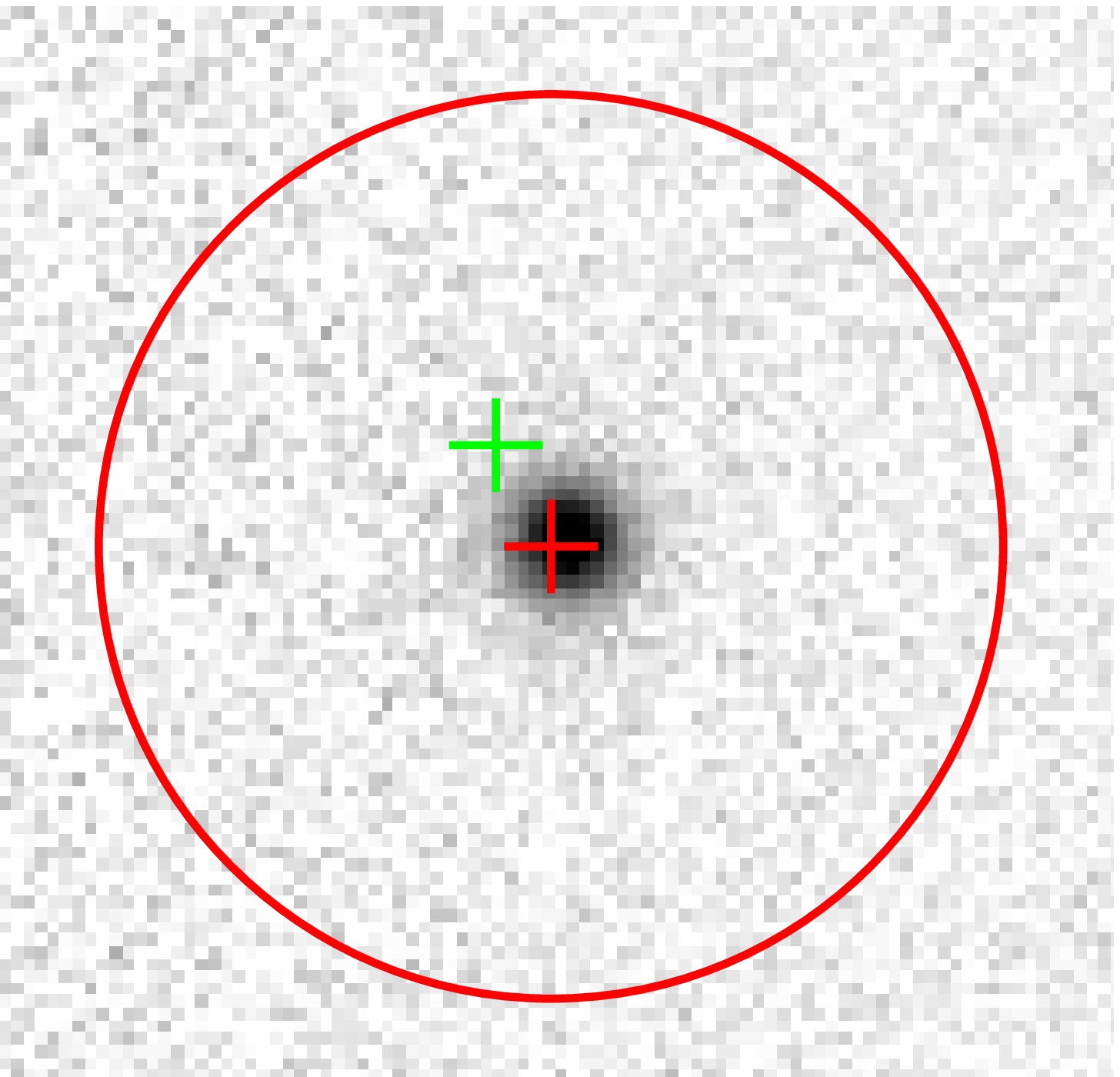}{0.25\textwidth}{(c)}
          \fig{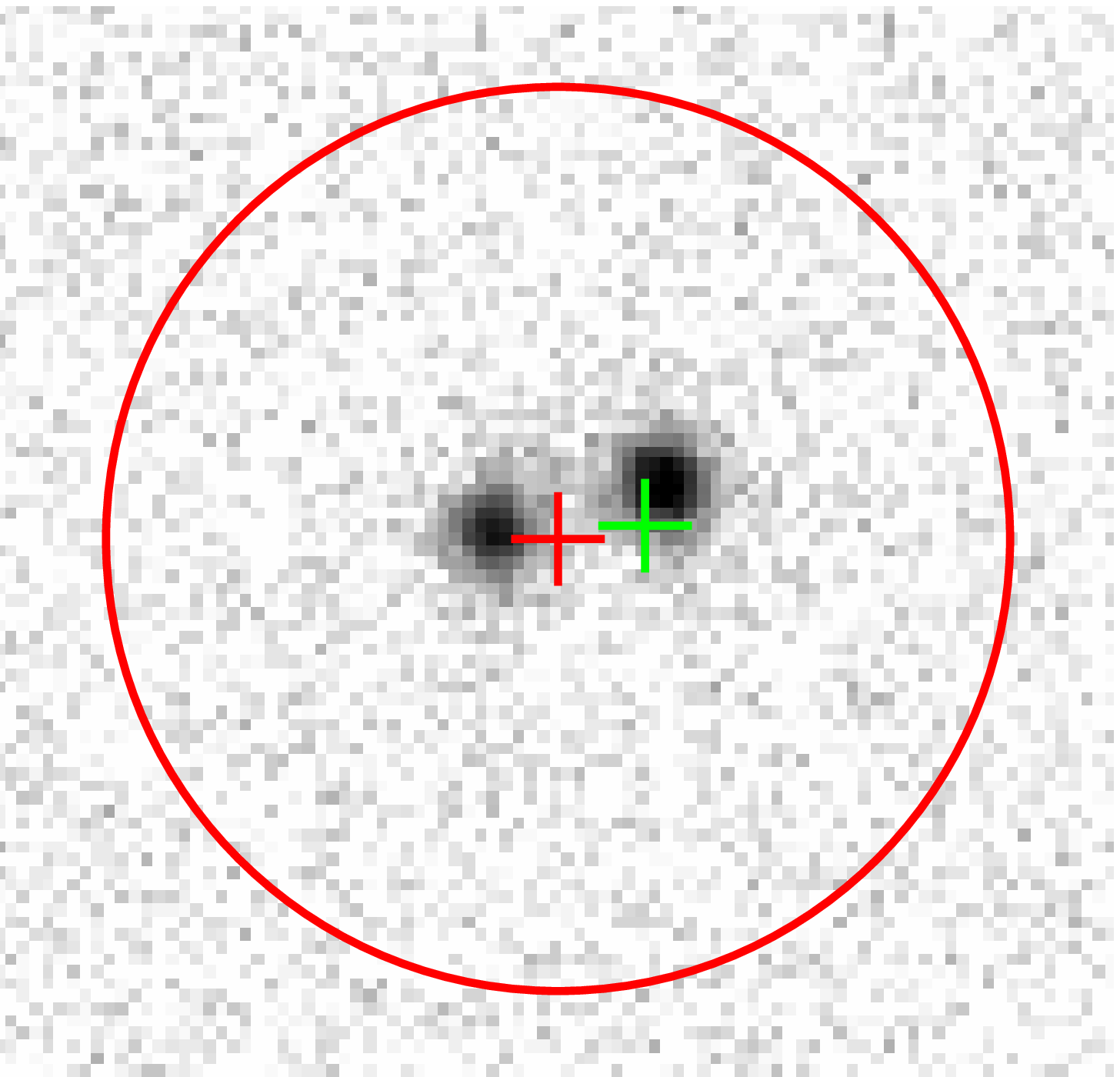}{0.25\textwidth}{(d)}
          }
\caption{Examples of survey images examined for nearby contaminants and the image quality flags they received, North up, East right. The green and red crosses show the EC coordinates (not corrected for proper motion) and All\emph{WISE} detection centers. The 7.8\arcsec radius red circle represents the limit of the ALL\emph{WISE} automatic source deblending routines which is capable of resolving sources at separations greater than 1.3$\times$FWHM of the \emph{W1} beam. Targets with additional sources within this circle are flagged as having potentially contaminated \emph{WISE} photometry.  From left to right: Vista VHS \emph{K$_{s}$} band images showing (a) EC 05267-4305 clean image, Im Flag: 00 (b) EC 05024-5705 potentially contaminated \emph{WISE} photometry, Im Flag: 01 and pair of VST ATLAS \emph{z} images showing (c) EC 21335-3637 clean image, Im Flag: 10 (d) EC 01107-1617 potentially contaminated \emph{WISE} photometry, Im Flag: 11}
\end{figure*}

In this study we focus on hydrogen atmosphere (DA) white dwarfs identified in the EC Survey. The EC Survey utilized \emph{U}-\emph{B} colors to select candidate objects and relied on follow-up photometry and low-resolution spectroscopy to classify each blue object \citep{sto97}. The authors follow the identification scheme described in \cite{sio83} to identify common white dwarf types, and note that the broad spectral features of white dwarfs make them easy to classify. Of the 2,637 unique hot objects identified in the EC survey, we find that 489 have been designated as type DA or possible type DA (e.g. DA?, DAweak, etc.). Candidates that have an uncertain but possible DA spectral type are also included in this study and are discussed later in the context of possible contaminants. 

For each target, we extract additional photometry from the \emph{GALEX} All-sky Imaging Survey GR5 (\cite{bia14}, hereafter \emph{GALEX}), AAVSO Photometric All-Sky Survey DR9 (\cite{hen16}, hereafter APASS), 2MASS All-Sky Point-Source Catalog \citep{cut03}, VISTA Hemisphere Survey (\cite{mcm13}, hereafter VHS), and the All\emph{WISE} Data Release of the Wide-Field Infrared Survey Explorer \citep{wri10}. Data collection for the EC Survey began in the 1980s so in order to minimize source mis-identification while cross-matching our targets across nearly three decades worth of surveys, we also collected proper motions for our targets from the PPMXL catalog \citep{ros10}. Using the J2000 epoch from PPMXL, we queried each photometric catalog for sources within 2.5\arcsec of the proper motion corrected target position, corrected to the mid-point of each survey's data collection period. 

To ensure we have selected the correct PPMXL source, we use a method similar to that described in \cite{gen17}, which essentially selects all nearby sources from the proper motion catalog of choice, then corrects their positions to the known epoch of the target before automatically selecting the nearest source. Unfortunately, the later releases of the EC Survey are increasingly lacking in epochs for each object coordinates, so the procedure was modified to enable user selected sources. To do this, we overlaid the J2000 corrected EC target coordinates and proper motion projections of all PPMXL sources within 15\arcsec on POSS2 imaging plates \citep{rei91}. This search returned 3 or less PPMXL sources for 468/489 candidates with many having only a single nearby source, leading to simple, unique source identifications based on proximity to the target coordinates. 4/489 targets had no PPMXL sources within 15\arcsec. For the remaining 17/489 targets where multiple sources were found near the target coordinates, we selected the PPMXL source most consistent with both the EC Survey position and the measured EC Survey B magnitude. These results suggest that up 5\% of our candidates could be mis-identified, potentially leading to spurious infrared excess selections or classifications, which we discuss in later sections.

It is well known that the large PSF of the \emph{WISE} beam ($\sim$\,6.0\arcsec\,in \emph{W1}) can lead to contamination from nearby sources, and care must be taken to ensure the measured All\emph{WISE} fluxes are consistent with a single source \citep{deb11,bar14}. To identify targets with potentially contaminated \emph{WISE} photometry, we collected cutouts of survey images from the VISTA-VHS and VST-ATLAS \citep{sha15} catalogs in \emph{K$_{s}$} and \emph{z} bands from the VISTA Science Archive. The VISTA Data Flow System pipeline processing and science archive are described in \cite{irw04}, \cite{ham08}, and \cite{cro12}. When both images were available, the VHS \emph{K$_{s}$} images were preferred as the photometric band is much closer to the \emph{WISE} photometry. Examples of images from each catalog are shown in Figure 1. Overplotted on each image is the position of the white dwarf target as identified in the EC Survey, and the position of the corresponding All\emph{WISE} detection, including a 7.8\arcsec\, circle around the \emph{WISE} position, which is the approximate limit of the automatic deblending routine used for the ALL\emph{WISE} pipeline. The imaging circle allowed us to quickly identify and flag targets with potentially contaminated \emph{WISE} photometry. Each target was assigned an image quality flag based on the results of studying the collected images by eye, which is included in the summary tables in the appendix. Targets that were identified as having potentially contaminated \emph{WISE} photometry were not excluded and should not be ruled out without more careful analysis of the contaminating source, but their \emph{WISE} excess should be given more scrutiny than those with clean images.

\section{White Dwarf Model Fitting and Infrared Excess Identification}

Our first step in identifying systems with an infrared excess is fitting the collected photometry of each target with a white dwarf model. We use a grid of hydrogen atmosphere white dwarf models, kindly extended to include \emph{GALEX} and \emph{WISE} photometry by P. Bergeron \citep{ber95}. To ensure the model and collected photometry were on the same magnitude scale, we applied zero-point offsets to the \emph{GALEX}, EC, APASS, 2MASS, and VHS magnitude\footnote{VHS magnitudes were transformed to the 2MASS system using the color-color equations described at http://casu.ast.cam.ac.uk/surveys-projects/vista/technical/photometric-properties} as defined in \cite{hol06} and \cite{cam14}. Each transformed magnitude was then converted to flux density units using published zero points \citep{coh03a,coh03b,jar11}. Our photometric uncertainties are derived from reported catalog values, and we assume a 5\% relative flux uncertainty floor. 

\begin{figure}
\epsscale{1.20}
\plotone{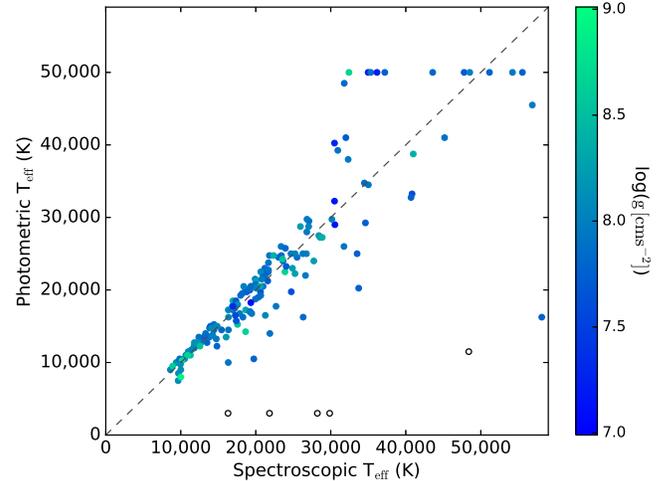}
\caption{Our photometric white dwarf temperatures assuming $\log {g}$=8.0 compared with spectroscopic fits from \cite{koe09} and \cite{gia11}. Open symbols represent targets with less than 3 datapoints used to constrain the photometric fit. The colorscale represents the spectroscopic surface gravity. A handful of white dwarfs can be seen hitting the top of our photometric grid at 50,000\,K.}
\end{figure}

Compared with previous WIRED surveys, our sample has a few unique features; first, our targets do not have a consistent set of optical measurements which are needed to anchor the white dwarf model photometry. The \emph{U}, \emph{B}, and \emph{V} band photometry from the EC Survey is incomplete and the APASS survey is ongoing, leading to sporadic coverage across the different photometric bands. The second is that fewer than half of our white dwarfs have a prior spectroscopic effective temperature and surface gravity determination, the latter of which is often necessary to split the degeneracy between the solid angle subtended by the white dwarf and its photometric distance. 

To address these issues, for our photometric white dwarf model fits we fixed the surface gravity of our model atmosphere grid to $\log {g}\,$=$\,$8.0 ($g$ measured in cm\,s$^{-2}$). For each model in our grid, we determine an initial flux scaling based only on the available optical data (0.4$\mu$m$\,\leq\,\lambda\,\leq\,$0.7$\mu$m). The photometric scale factors were then transformed to initial distance estimates, and following the prescription of \cite{har06}, we apply photometric reddening corrections to our photometry for all sources beyond 100pc. The white dwarf model was then re-fit to the corrected photometry. The best-fitted model was chosen by minimizing the chi-square metric as computed for each scaled model using all photometry at wavelengths below 1.0 $\mu$m. As discussed in Section 4.2, for white dwarfs identified with strong stellar excesses that obviously extended into the near-infrared and optical, we limited the photometry used to determine the best-fit to wavelengths below 0.5 $\mu$m. For those targets with prior spectroscopic $\log {g}$ and T$_{\rm eff}$ solutions from either \cite{koe09} or \cite{gia11}, we assumed the spectroscopic $\log {g}$ and T$_{\rm eff}$ for our white dwarf atmospheric parameters, and generated model photometry scaled to the observed, de-reddened photometry with the method described above.

We compare our photometrically derived effective temperatures to the spectroscopic determinations for apparently single objects where we have both in Figure 2, with the spectroscopic surface gravity displayed as a colorscale. The most egregious outliers on Figure 2, shown as open symbols, are cases where fewer than 3 photometric points were available to constrain the photometric fit. We note that none of our new infrared excess candidates suffered from this severe lack of data. Using the scatter in the relationship, we can establish uncertainties for our white dwarfs which only have photometric fits. Below 15,000\,K, the fits are generally good, with an uncertainty of $\sim$\,1000\,K. Between 15,000--30,000\,K, the scatter is greater, resulting in an uncertainty of $\sim$\,3000\,K. Above 30,000\,K, the photometric fits are generally unreliable which reflects a lack of short wavelength optical and ultra-violet photometry needed to constrain the bluer SEDs. Despite the agreement below 30,000\,K, there are still a handful of 3$\sigma$ outliers given the uncertainties above, all of which exhibit a bias toward lower photometric temperatures. We find the culprit to be sporadically poor \emph{U} band photometry from the EC Survey, examples of which can be seen in the SEDs of EC 02566-1802 and EC 23379-3725 in Figures 4 and 9 respectively. Unfortunately, we found no way to determine a priori if the EC \emph{U} band photometry was poor, and therefore cannot correct for it in this sample. Based on the number of 3$\sigma$ outliers in Figure 2 below 30,000\,K we estimate the poor \emph{U} band photometry to be affecting less than 10\% of our sample.

Targets showing a 5$\sigma$ excess in either the \emph{W1} or \emph{W2} bands or a 3$\sigma$ excess in the \emph{W1} and \emph{W2} bands were flagged as infrared excess candidates. These criteria flagged 111 out of 378 white dwarfs with All\emph{WISE} detections as infrared excess candidates. These candidates comprise the sample discussed in the remainder of the paper. 

\section{Infrared Excess Classification}

While the selection of infrared excess candidates is straightforward, classification of the infrared source without a clean separation of the SEDs can be misleading. For programmatic searches of infrared excesses like the prior WIRED studies \citep{deb11,hoa13}, the root of this problem is the shortage of infrared excess data points, which in our case is exacerbated by an incomplete near-infrared dataset. Techniques for classifying excesses as dusty debris disks are particularly lacking. Conventional color-color selection \citep{hoa13} can miss the subtle infrared excesses that likely comprise a majority of dusty debris disks \citep{roc15,bon17}, and it does not make use of all of the available information gained from fitting a model white dwarf atmosphere to the observed photometry, namely the photometric distance. More complex models can in theory distinguish between stellar and dusty infrared excesses \citep{deb11}, but reduced chi-square metrics are often degenerate between stellar and dusty classifications given the limited number of infrared excess points constraining the models. Because of these concerns, and our uniquely deficient photometry, we sought a simple technique that could distinguish between whether the excess is consistent with a dusty debris disk, an unresolved stellar or sub-stellar companion, or a background contaminant, and that would quickly highlight the best candidates for follow-up studies.  

\begin{figure*}[ht]
\plotone{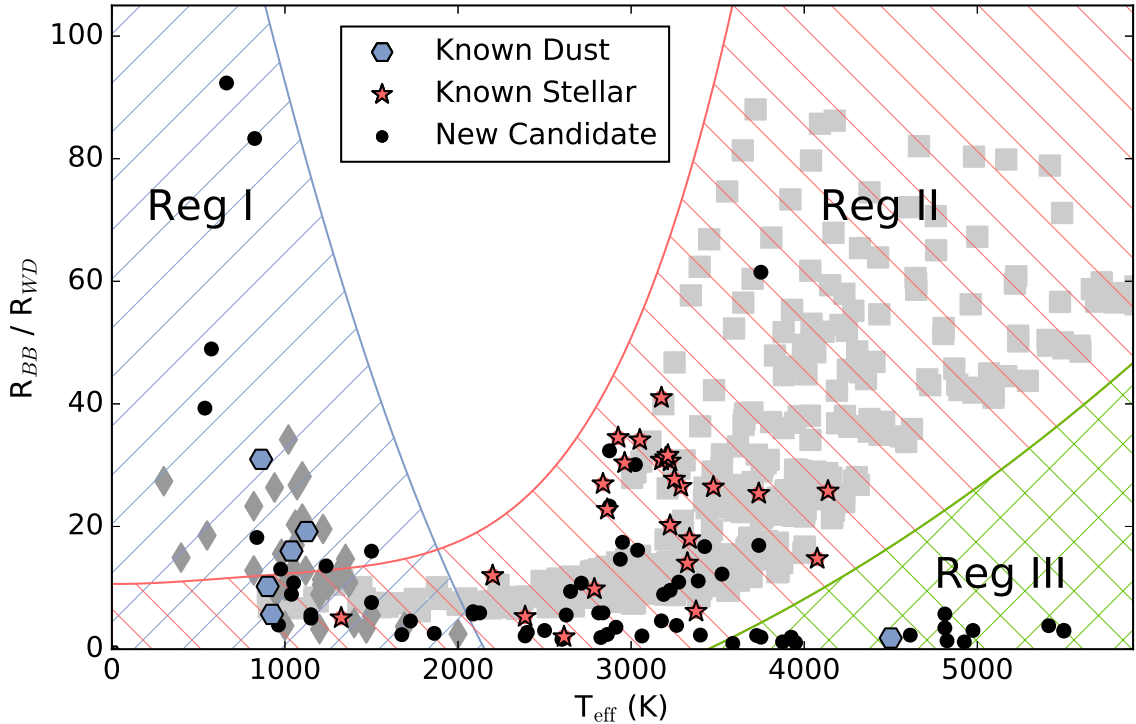}
\caption{Plot of best-fitting blackbody T$_{\rm eff}$ and radius for each infrared-excess candidate (black circles), with the blackbody radius scaled to the white dwarf radius. Literature identified WD+dM systems are denoted with red stars while identified dust excesses are denoted with blue hexagons. Light grey squares show stellar and sub-stellar model effective temperatures and radii from the models of \cite{cha97} and \cite{cha00} as scaled to a $\log {g}=8.0$ white dwarf radius. Grey diamonds show the effective temperatures and radii adopted from the single temperature blackbody fits of all \emph{Spitzer} confirmed dusty debris disks by \cite{roc15}.}
\end{figure*}

The simplest model that describes the infrared excess is a single temperature blackbody assumed to be at the photometric distance of the white dwarf star. Assuming the white dwarf atmospheric parameters and photometric distances derived above, we fit a single temperature blackbody to the observed infrared excess for each infrared excess candidate in our sample, with only the blackbody effective temperature and radius as free parameters. Figure 3 shows the results of the single temperature blackbody fits for our entire sample, plotted as the effective temperature versus radius of the infrared source as scaled to the white dwarf radius. It is important to keep in mind that what is actually being fitted is a ratio of the solid angle subtended by the single temperature blackbody source to that of the white dwarf star, and so errors in the assumed white dwarf radius and distance, particularly for those with photometric atmospheric solutions where we have assumed a surface gravity, propagate into this measurement. Figure 3 should not be interpreted as giving an accurate description of the temperature and radius of the infrared source, particularly for dusty debris disk candidates which are neither perfectly circular nor at a single temperature, but rather it can be used as guidance for selecting targets of interest. To help guide the reader, we also plot in the background as light grey squares the effective temperature and radius for the stellar and sub-stellar models of \cite{cha97} and \cite{cha00}, which extend from early M dwarf stars down to sub-stellar and late type brown-dwarf stars, scaled to a typical $\log {g}=8.0$ white dwarf radius. Finally, we also show the parameters for single temperature blackbody fits to all known dusty debris disks confirmed with \emph{Spitzer}, independently fitted by \cite{roc15}, in the background as grey diamonds. We have derived the blackbody radius used in the fits of \cite{roc15} from the white dwarf effective temperature, blackbody temperature, and fractional infrared luminosity from their Table 3, assuming the radius of a $\log {g}$=8.0 white dwarf, consistent with what the authors used when fitting their single temperature blackbodies.

We also performed a literature search for all of the objects in our sample and found 6 with published infrared excesses identified as dusty debris disks, and 44 with published infrared excesses identified as stellar or sub-stellar companions. We indicate these on Figure 3 with blue hexagons (dusty disks) and red stars (stellar companions) and see that, in general, these groups occupy distinct regions of the plot. We define three regions of interest in this plot. Region I is defined as a region of low effective temperature (T\,$<$\,2000$\,$K) and  varying radius. We see that the 5/6 known dusty debris disks in our sample cluster in this region in the lower left corner of our plot. The one identified dusty debris disk in our sample which does not follow this trend, PG 1457-086, is discussed in more detail in section 4.3. While the 5 literature-identified dusty debris disks in this region provide a nice set of boundaries for selecting new dusty debris disk candidates within our sample, the single temperature blackbody fits of the sample of 35 \emph{Spitzer} confirmed dusty white dwarf systems from \cite{roc15}, overplotted as grey diamonds, provide an independent view of the extent of the dusty debris disk region. In Region II, known stellar companions to white dwarfs congregate at higher temperatures and radii, in a locus around 3000$\,$K and 30$\,$R$_{\rm WD}$ (0.4$\,$R$_\odot$), which are the temperature and radii expected for an unresolved M dwarf type companion. Region II extends down into the low temperature and small radius regime of Region I, where the overlap between dusty debris disks and late-type stellar and sub-stellar companions forces us to a less certain conclusion about the source of the infrared excess. Objects in Region III, which consists of infrared excess best reproduced by objects of higher temperature (T\,$>$\,3500$\,$K) and small radius R\,$<$\,10\,R$_{\rm WD}$ (0.1$\,$R$_\odot$), have no obvious source, but are likely the result of mis-classification or contamination. We discuss each region and the objects they contain below. 

\begin{figure*}[t]
\includegraphics[keepaspectratio=True,width=1.0\textwidth]{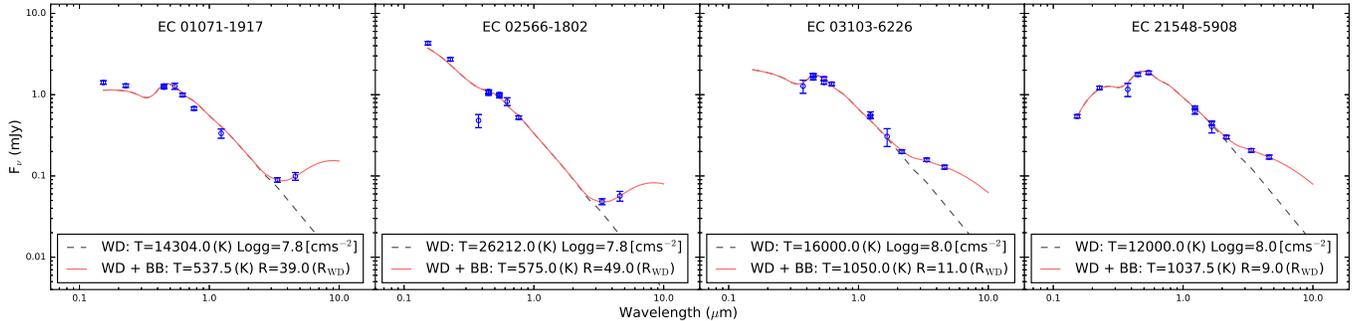}
\caption{SEDs of four of the most promising dusty debris disk candidates with white dwarf model and single temperature blackbody fits}
\end{figure*}

\subsection{Region I: Compact, Dusty Debris Disks}

In addition to the empirical boundaries given by the single-temperature blackbody fits to the known dusty debris disks, there is a corresponding theoretical expectation that dusty debris disks should congregate in this region. The formation of dusty debris disks via the tidal disruption of asteroids suggests they should not extend well beyond the asteroid tidal disruption radius at $~$1.0$\,$R$_\odot$, or $~$85$\,$R$_{\rm WD}$ for typical white dwarf masses around 0.6\,M$_{\odot}$ and asteroid densities $\sim$\,2\,g/cm$^{-3}$ \citep{ver14}. At their inner edge, the cm to micron sized dust is only expected to be able to survive at temperatures below 2000\,K before sublimating into gas \citep{raf12}. Since the dust within this region is expected be optically thick, the majority of it is shielded from direct radiation and it's temperature falls off rapidly with distance from the white dwarf \citep{chi97,jur03}, with the outer dust near the tidal disruption radius radiating at only a few hundred degrees kelvin. The temperature of the single-temperature blackbody fit along the x-axis of Figure 3 can be thought of as an area weighted temperature average of the dust disk, which given the expected inner and outer boundaries should be between 500-1500$\,$K depending on the width of the disk. The radius of the single-temperature blackbody plotted along the y-axis is less straightforward to interpret as it is dominated by the inclination of the dust disk. 

There is some overlap with the stellar candidates defined by Region II and the bottom of Region I. We choose to list the objects in this overlapping region as dusty debris disk candidates. Because the new dusty debris disk candidates are of particular interest to this study, we provide the spectral energy distributions with the single temperature blackbody fits for each new dusty debris disk candidate in Figure 4 and in the appendix in Figure 9. Table 1 provides a summary of the properties  of all objects in Region I, including new debris disk candidates, known debris disk systems recovered by our search, and candidates we have chosen to reject as dusty debris disks based on either an obviously poorly fit SED or an independent physical reason, detailed in the Rejected Candidates section of the appendix. We also performed extensive literature searches on each object in this region and provide notes, additional data, and recommendations for follow-up on each object. 

In the remainder of this section, we highlight our follow-up on four promising candidates which appear firmly inside of Region I, have high resolution spatial follow-up which suggests their \emph{WISE} photometry is unlikely to be contaminated, and have optical spectroscopy to search for atmospheric metal pollution. The signature of recent accretion is not necessarily independent confirmation that an observed \emph{WISE} infrared excess is consistent with a dusty debris disk. Recent studies have shown that the frequency of atmospheric pollution consistent with metal rich exoplanetary debris accretion could be as high as 25\% \citep{koe14}. However, to date every confirmed dusty debris disk hosting white dwarf has been demonstrated to be actively accreting metal rich debris \citep{far16} so limits on the atmospheric pollution from optical spectra can be used as an easy way to prioritize dusty debris disk candidates. The most commonly detected transition in the optical is the \ion{Ca}{2} K resonance line at 3934$\,$\AA \citep{zuc03,koe05,kaw11}. For each object we provide measurements or conservative upper limits to equivalent widths of the \ion{Ca}{2} K line based on archival or collected spectra. Ultimately, only higher spatial resolution, longer wavelength observations, or \deleted{low-resolution}near-infrared spectroscopy can confirm \emph{WISE} excesses as necessarily due to dust.
 
\begin{figure}[h!]
\epsscale{1.1}
\plotone{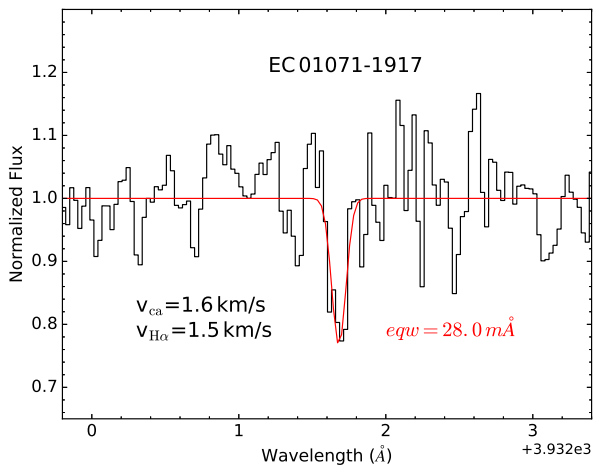}
\caption{Combined UVES SPY spectrum for EC 01071-1917, centered on Ca K. Equivalent width measurement shown in red.}
\end{figure}

\emph{EC 01071-1917:}
Also known as GD 685 and WD 0107-192. The best-fitted single temperature blackbody parameters place it well away from the overlapping stellar models of Region II. The ATLAS \emph{z} band image is free of nearby contaminants. This object was also targeted as part of the SPY survey \citep{koe01}, from which we have adopted the spectroscopic effective temperature and surface gravity (T$_{\rm eff}$=14,304 K, $\log {g}$=7.8) for our white dwarf model parameters in the SED fit. We collected and combined the pipeline reduced archival SPY spectra using the ESO Science Archive to search for atmospheric Ca. We detect a subtle Ca K absorption feature shown in Figure 5 with an equivalent width of 28.0$\,\pm\,9.0\,$m\AA\, at a heliocentric corrected velocity of 1.6$\,\pm\,$1.4$\,$km/s, which is consistent with the white dwarf photospheric velocity of 1.5$\,\pm\,$0.9$\,$km/s as measured with a gaussian fit to hydrogen alpha NLTE line core. 

\emph{EC 02566-1802:}
Also known as HE 0256-1802 and WD 0256-180. The best-fitted single temperature blackbody parameters place it well away from the overlapping stellar models. ATLAS \emph{z} band imaging is free of nearby contaminants. We also obtained additional \emph{K$_{s}$} follow-up with the SPARTAN infrared camera on the SOAR telescope which confirm the lack of nearby sources, but are unable to provide additional calibrated photometry due to poor observing conditions. We adopt a spectroscopic temperature of 26,120\,K and surface gravity of $\log {g}$=7.76 from the SPY survey \citep{koe09} for our SED fits, which agrees with our independent photometric temperature. The white dwarf effective temperature is high, but not prohibitively so, for white dwarfs which host dusty debris disks. At this temperature, the optical spectra are less useful as probes for atmospheric pollution due to higher ionization states of atmospheric metals with transition wavelengths in the ultra-violet \citep{koe05, koe14}. Nevertheless we place an upper limit to the eqw \ion{Ca}{2} K line of 17.5$\,$m\AA. A nearby absorption feature of $\sim$\,21.0$\,$m\AA\, was detected at 10.3$\,\pm\,$1.1$\,$km/s, but is inconsistent with the white dwarf photospheric velocity of 25.5$\,\pm\,$1.7$\,$km/s, and is likely interstellar. 

\emph{EC 03103-6226:}
Also known as WD 0310-624. The white dwarf model is well constrained and the best-fitted single temperature blackbody parameters place it within the region of known dusty debris disks, though there is some overlap with the stellar models of Region II. VHS \emph{K$_{s}$} band imaging is free of nearby contaminants. This object was identified as a white dwarf candidate and followed up spectroscopically by \cite{sub07}, where it is noted that the difference in spectroscopic and photometric temperatures are suggestive of it being an unresolved double-degenerate candidate. Our independent photometric temperature is slightly hotter than their photometric fit (15,250\,K vs 13,900$\,$K) and in better agreement with their spectroscopic temperature of $\sim$\,17,000$\,$K. Considering that both estimates of the photometric temperature assume a surface gravity of $\log {g}$=8.0, the remaining discrepancy could be within the photometric temperature error budget. We also obtained high signal-to-noise optical spectroscopy on 2015 November 18 with the Goodman Spectrograph on SOAR. We used the 0.46\arcsec slit in combination with the 1800\,l/mm grating to cover a wavelength range from 3740\,\AA\, to 4580\,\AA\, with a resolving power of R\,$\sim$\,7000. No absorption features are detected near Ca K, resulting in an upper limit of 46.5$\,$m\AA. 

\emph{EC 21548-5908}
The white dwarf model is well constrained by multiple near-infrared photometry points (2MASS and VHS). VHS \emph{K$_{s}$} band imaging is free of nearby contaminants. To confirm the atmospheric parameters, we obtained low resolution optical spectroscopy with the Goodman Spectrograph on SOAR covering the hydrogen Balmer series from H$\beta$ blueward. Using the techniques described in \cite{fuc17}, we fit the spectra to a grid of hydrogen atmosphere white dwarf models kindly provided by D. Koester \citep{koe10}, and determined a spectroscopic effective temperature and surface gravity of 12,330\,K and $\log {g}$=8.04, which are consistent with the photometric fit. We also obtained high signal-to-noise optical spectroscopy on 2015 October 18 with the Goodman Spectrograph on SOAR, using the same instrument setup described for EC 03103-6226. No absorption features are detected near Ca K, resulting in an upper limit of 65.9$\,$m\AA. 

\subsection{Region II: Unresolved Stellar/Sub-Stellar Companions}

The results of objects that fall within Region II are summarized in Table 2. We provide examples of SEDs of objects with the stellar classification in Figure 11 of the . We note that for the brighter stellar excesses which begin in the optical wavebands, our white dwarf models often attempted to fit some of the additional optical flux from the unresolved companion, resulting in white dwarf model fits that were systematically overluminous, and single temperature blackbody fits that were systematically underluminous, pulling the single temperature blackbody fits to lower temperatures. To account for this, we re-fit the white dwarf models restricting our photometry to wavelengths less than 0.5 $\mu$m. The positions of objects in Figure 3 and the fitted parameters given in Table 2 reflect our best-fitted values after this correction was applied. More detailed modeling is necessary to determine the stellar companion spectral type, and is beyond the scope of this paper. 

\subsection{Region III: High Tempertaure, Small Radius}

The third region of Figure 3 consists of objects with an infrared excess that is best fitted by a single temperature blackbody with a high temperature (T$\,>\,$4000$\,$K) and small radius (R$\,<\,$10$\,$R$_{\rm WD}$). The results of objects that fall within this region are summarized in Table 3. Examples of objects in this region are shown in the appendix in Figure 12. The nature of the objects populating this region is less obvious than the other two regions. Furthermore, since Region III includes one of the known dusty debris disks in our sample, PG 1457-086, we took great care in investigating this region. We propose these objects are the result of some combination of the following four scenarios.

1) Erroneous stellar classification/poor photometry/poor WD model fit: The first thing that stands out when looking through the results in Table 3 is that 6/15 of the objects have an uncertain EC spectral type and 5 of them are identified as potential hot sub-dwarfs. With uncertainty about the blue object classification, we can no longer rely on our DA white dwarf atmospheric models to accurately predict the photometric flux in the near-infrared. 5/9 remaining objects have high signal-to-noise spectral follow-up and atmospheric model parameter fits from either the Gianninas or Koester spectroscopic surveys so they cannot be accounted for by misclassification. 

Finally, if the white dwarf model is fitted at a higher temperature than the true white dwarf temperature, the difference in slope can result in the excesses observed in this section. This appears to be the case for at least one of the white dwarfs where we have assumed a spectroscopic temperature, EC 10188-1019, as evidenced by the highly discrepant \emph{GALEX} photometry. Our fitted photometric temperature is much lower than the reported spectroscopic temperature from \cite{gia11} (11,000\,K vs 17,720\,K), and the difference in temperature completely accounts for the observed infrared excess.  We note that EC 10188-1019 is flagged by \cite{gia11} as being magnetic, which is likely affecting the spectroscopic temperature as non-magnetic models were assumed for the fits. 

As discussed in both \cite{koe09} and \cite{gia11}, the Balmer features used to determine spectroscopic atmospheric parameters peak in strength around 13,000-14,000\,K, and fits of white dwarfs near this temperature often suffer from a hot/cold solution degeneracy across this boundary. Photometric fits are one way to break this degeneracy. 3 white dwarfs in this region have spectroscopic temperatures near this boundary, the magnetic white dwarf EC 10188-1019 discussed above, EC02121-5743, and EC 22185-2706. Cooler white dwarf models could explain the observed excess.

2) Irradiated sub-stellar/planetary mass companion: The best-fitted radii of objects in this region are consistent with Jupiter and brown-dwarf sized companions, but the temperatures require a substantial amount of additional heating. The post-main sequence evolution of the white dwarf progenitor is expected to result in planetary re-heating via accretion and irradiation \citep{spi12}, but the temperatures needed to fit the infrared excesses in this region (T$\,>\,$4000$\,$K) are well beyond what is expected. Furthermore, the thermal relaxation timescale of the re-heated planets is on the order of hundreds of millions of years, meaning the re-heated planets would only be expected around the youngest white dwarfs in this sample \citep{spi12}. 

There are a handful of confirmed white dwarf-brown dwarf binaries in compact orbits which suggest that, despite engulfment, the brown-dwarf survives post-main sequence evolution relatively unscathed \citep{far04,max06,cas12,ste13}. The compact orbits lead to tidally synchronous orbits and significant differences between dayside and nightside brown-dwarf surface temperatures \citep{cas15}. Despite the strong irradiation, the dayside temperatures ($\sim$\,3000$\,$K) are still too cool to explain the excesses seen in this region \citep{cas15}. Nonetheless, if these excesses are the result of irradiated brown-dwarf companions, there should be observable spectral features in the near-infrared and optical, and brightness modulations from tidal and reflection effects from the companion. The lack of these features could quickly rule out compact brown-dwarf companions. 

Finally, it is also worth pointing out that previous studies have found the brown dwarf companion fraction to white dwarfs to be low. In a search for binary companions that included near-infrared direct imaging and near-infrared excess techniques, \cite{far05} find the white dwarf brown dwarf companion fraction to be $<$ 0.5\%. In a previous WIRED study, \cite{deb11} find the observed frequency to be between 1.3$\pm$0.6\% after accounting for likely contaminants. Even in the most optimistic case, in our sample of 383 white dwarfs with \emph{WISE} detections we should only expected 7-8 white dwarf brown dwarf systems, which is not sufficient to explain all of the observed excesses in this region.

3) Unresolved contaminants: Another way to produce the subtle excess is with an unresolved, line-of-sight object which is at a different distance than the white dwarf. Even with high quality near-infrared follow-up from the VISTA-VHS survey, care must still be taken to confirm the excesses seen in this section are not the result of a contaminant. We find EC 14572-0837, also known as PG 1457-086 (hereafter EC 14572), to be an example of contamination by an unresolved background source.

EC 14572 is identified in the literature as a dusty debris hosting white dwarf, with strong atmospheric metal pollution \citep{far09}. The infrared excess was confirmed by \emph{Spitzer} and determined to be most consistent with a dusty debris disk. EC 14572 is included in the sample of \emph{Spitzer} confirmed dusty debris disks identified by \cite{roc15}

\begin{figure}[h!]
\epsscale{1.18}
\plotone{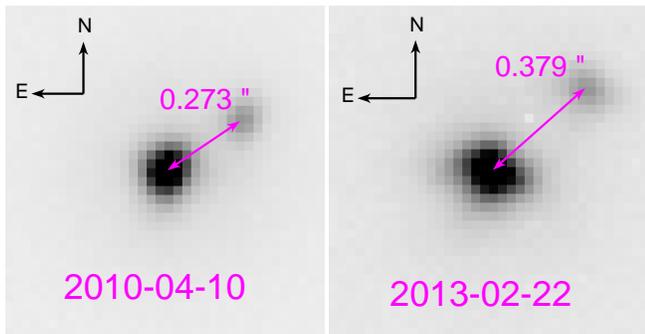}
\caption{Two epochs of NAOS+CONICA high resolution near-infrared images of EC 14572. The two \emph{J} band images show a very nearby source which does not appear to be in a common proper motion pair with the white dwarf.}
\end{figure}

EC 14572 is the only published dusty debris disk that we failed to correctly identify in our single temperature blackbody selection. Because this might indicate a short-coming in our dusty debris disk selection technique, we performed a thorough archival data search to determine if any additional data could help resolve the discrepancy between the infrared excess as seen when scaling to the optical versus near-infrared photometry. Our ESO archive search revealed that EC 14572 was a target in a multi-epoch, high-contrast and high-spatial resolution imaging search for Jupiter sized planets around dusty debris disk hosting white dwarfs with the NAOS+CONICA near-infrared imager on the VLT as a part of program 085.D-0673(A) led by M. Radiszcz. The high quality \emph{J} band imaging reveals a close contaminant. Figure 6 presents two epochs of imaging from this study for EC 14572, taken 3 years apart, centered on the brighter object, with approximate separation measurements. 

The change in separation between the two epochs of 0.126\arcsec\, is consistent with the direction and magnitude of the proper motion of EC 14572 as measured by the PPMXL survey ($\mu_\alpha=$2.4\,mas\,yr$^{-1}$ and $\mu_\delta$=-38.8\,mas\,yr$^{-1}$), suggesting the two objects are not in a common proper motion pair. We performed aperture photometry on the two sources and find the flux ratio between the white dwarf and the contaminant to be 3.2$\pm$0.6, which is agreement with the excess flux above the white dwarf model observed in the \emph{J} band. The best-fitted blackbody that can explain the excess has a temperature of 4400\,K.

Any additional flux from the unforeseen companion is certainly contaminating our near-infrared and \emph{WISE} data, and is very likely present in the \emph{Spitzer} data of \cite{far09}. Though \cite{far09} scaled the white dwarf model photometry to the near-infrared data when modeling the excess as a dusty debris disk, effectively including the near-infrared flux of the unresolved contaminant in their stellar model, if the contaminant is significantly cooler than the white dwarf the difference in slope of the spectral energy distribution at these wavelengths could be solely responsible for the infrared excess measured for EC 14572. Given the difference in proper motion between the white dwarf and contaminant, increased separation should allow future follow-up to independently measure the near-infrared flux of the white dwarf and the contaminant, and definitively resolve the source of the excess infrared radiation.

\section{Locus of Gaseous Debris Hosting Disks}
As a by-product to our search for new dusty debris disks, we also found an interesting relation among dusty debris disks which are also known to host gaseous debris in emission. Circumstellar gaseous emission has been observed in the optical spectra of 8 white dwarfs which also dusty debris disks \citep{man16b}. The gaseous debris is believed to be spatially coincident with the dusty debris \citep{mel10}, and the interaction between the gas and dust is likely to play large role in the evolution and accretion of the dust disk \citep{met12}. The double-peaked emission calcium triplet emission features exhibited by these disks lend themselves to more detailed dynamical modeling than can be accomplished with the infrared excesses of dusty debris disks \citep{gan06,har16}, and several have been shown to be variable on timescales of decades \citep{man16b}. 

\begin{figure}[h]
\epsscale{1.15}
\plotone{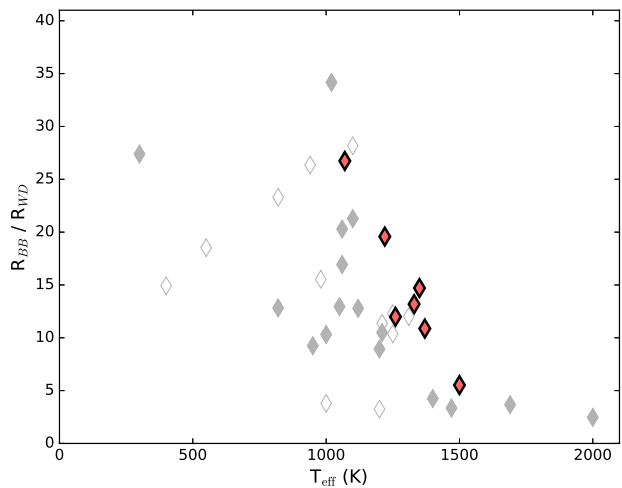}
\caption{Location of all \emph{Spitzer} confirmed dust disks as fitted with single temperature blackbodies by \cite{roc15}. Systems with observed Ca triplet emission are identified with red diamonds with black outlines, and systems where we found no evidence of Ca triplet emissions with our SOAR observations are shown as filled grey diamonds. Open diamonds are systems we have not yet followed-up for Ca triplet emission.}
\end{figure}

Figure 7 shows an expanded region of our Figure 3 with the blackbody fits from \cite{roc15}. It is worth reiterating here that the disks themselves are not spherical, and the best fitted ``radius'' is a proxy for apparent surface area, which is affected by inclination and the inner/outer disk radius. In red diamonds with black outlines we highlight all of the literature identified gaseous debris systems within the \cite{roc15} sample. We have also been surveying known dusty debris disk hosting white dwarfs for calcium triplet emission, the most common tracer of gaseous debris in these systems, and we include our non-detections as filled grey diamonds. Our Ca triplet observations are being carried out with the Goodman Spectrograph \citep{cle04} on the SOAR telescope, using the 1200\,l/mm grating and the 1.07\arcsec slit with a wavelength coverage of 7900\,\AA\, to 9000\,\AA\, and routinely reach a signal-to-noise of $\sim$\,40 per 1.2\AA\, resolution element. Examples of non-detections are shown in Figure 8. Open grey diamonds are systems we have not yet surveyed. 

The white dwarfs with gaseous and dusty debris all lie along the terminus of the dusty white dwarf region of the blackbody fit plot. In other words, for any projected surface area, the dusty disks with gaseous debris congregate at the highest temperatures. This is not surprising as the high temperature side of the dust disk region should be defined by the dust sublimation temperature. It seems natural then that gas disks appear most frequently in systems with copious amounts of dust at the sublimation temperature.  

\begin{figure}[h]
\epsscale{1.15}
\plotone{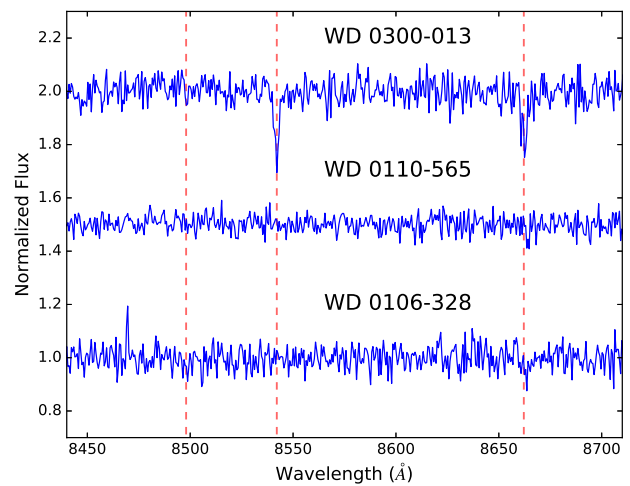}
\caption{SOAR spectra used to rule out Calcium triplet emission in three dusty debris hosting white dwarfs. The dashed red line shows the rest wavelength of the calcium triplet. Atmospheric Calcium is seen in absorption in the DB white dwarf WD 0300-013.}
\end{figure}

Another interpretation is that for a given temperature the white dwarfs with gaseous debris host the largest, and therefore most luminous dusty debris disks. The observation that white dwarfs with gaseous and dusty debris tend to have brighter dusty debris disks is not novel \citep{far16}, but we find it particularly interesting in the context of the results from the dynamical modeling that has been performed on the emission profiles. Typically, high inclinations ($i\,>\,60^\circ$) are needed to reproduce the large velocity dispersion and deep inner regions of the Ca emission profiles \citep{gan06,gan07,mel12}. This is difficult to reconcile with the brightness of the infrared dusty components, as all other things considered equal, one would expect the low inclination, face-on dust disks to be the highest luminosity disks. The implication is that systems which host gaseous debris in combination with dusty debris may not display equivalent, flat geometry as those without gaseous debris. This could be expected if the gas was collisionally produced, perhaps during a recent disruption or collision with an existing disk as described in \cite{jur08}. 

\section{Conclusions}
The EC Survey has provided a number of new, bright, spectroscopically confirmed white dwarf stars in the Southern Hemisphere, which we have surveyed for infrared excesses. The challenges of extending the WIRED techniques to a survey with incomplete spectroscopic and photometry were discussed, and a new technique for separating dusty debris disk candidates from stellar companion candidates based on single temperature blackbody fits to the excess radiation, which yields four new promising dusty debris disk candidates. We emphasize however that all infrared excesses discussed in this paper should be considered as candidates until independently confirmed. The selection of dusty debris candidates via single temperature blackbody fits works in a uniform way with good to poor photometry, and should prove useful for \emph{Gaia} white dwarf infrared excess studies. \emph{Gaia} searches will benefit greatly from the independent distance estimates and the precise, space-based \emph{G} band flux measurement, which can be used to anchor the white dwarf model photometry.

Along the way, we identified EC 14572 as an outlier among the literature identified dusty debris disks, and an archival search reveals high-contrast, high-spatial resolution imaging that suggests the observed excess could be contaminated by an unresolved contaminant. It remains to be seen if the companion can account for all of the observed infrared excess, or if the system still requires a dusty debris disk. We also identified the gaseous debris disk hosting white dwarfs on the blackbody temperature and radius plane, and find that they form the terminus for dusty debris disks, providing clues to their origin and evolution.

\acknowledgments
We would like to thank the anonymous referee for a detailed review which greatly improved this manuscript. E. Dennihy, J. C. Clemens, P. C. O'Brien, and J. T. Fuchs acknowledge the support of the National Science Foundation, under award AST-1413001. D. Kilkenny acknowledges financial support from the National Research Foundation of South Africa. This work is based on data obtained from (1) the Wide-Field Infrared Survey Explorer, which is a joint project of the University of California, Los Angeles, and the Jet Propulsion Laboratory (JPL), California Institute of Technology (Caltech), funded by the National Aeronautics and Space Administration (NASA); (2) the Two Micron All Sky Survey, a joint project of the University of Massachusetts and the Infrared Processing and Analysis Center (IPAC)/Caltech, funded by NASA and the National Science Foundation (NSF); (3) the ESO Science Archive Facility (4) the Southern Astrophysical Research (SOAR) telescope, which is a joint project of the Minist\'erio da Ci\^encia, Tecnologia, e Inova\c{c}\~{a}o (MCTI) da Rep\'ublica Federativa do Brasil, the U.S. National Optical Astronomy Observatory (NOAO), the University of North Carolina at Chapel Hill (UNC), and Michigan State University (MSU); (5) the VizieR catalog access tool, CDS, Strasbourg, France; (6) the NASA/IPAC Infrared Science Archive, which is operated by JPL, Caltech, under a contract with NASA; (7) the NASA Astrophysics Data System. This research has made use of the SIMBAD database, operated at CDS, Strasbourg, France.

\newpage
\appendix
\section{Notes on Remaining Dusty Debris Disk Candidates}
\emph{EC 00169-2205:}
Also known as GD 597 and WD 0016-220. This object has a tenuous excess, which does not continue into the \emph{W2} band, although the error bars on \emph{W2} are large. It was included as part of a high resolution imaging survey by \cite{far05} to uncover low luminosity companions to white dwarfs and no companion was detected. EC 00169-2205 was also included in the \cite{zuc03} search for metals in white dwarfs via the \ion{ca}{2} K line. No Ca was detected, and \cite{zuc03} provides an upper limit to the Ca K equivalent width of $<\,$10\,$\AA$, which at this temperature, corresponds to an atmospheric abundance upper limits of [Ca/H]$\,<\,$-10. This would be an unusually low abundance for an object with an infrared bright dust disk \citep{far12} and we therefore believe the detected excess to be the result of poor \emph{WISE} photometry. \\

\emph{EC 01129-5223:}
Also known as JL 237. This object has a tenuous excess, which does not continue into the \emph{W2} band, although the error bars on \emph{W2} are quite large and the excess does begin in the Ks band which is high quality. The VHS \emph{K$_{s}$} band image available is free of nearby contaminants. The spectral energy distribution is not very well constrained by the limited optical photometry, but the departure from blackbody in the \emph{K$_{s}$} band and \emph{W1} band are statistically significant.  \\

\emph{EC 05276-4305:}
The excess is small but the white dwarf model is well constrained by the multiple near-infrared data points (2MASS and VHS). The VHS \emph{K$_{s}$} band image is free of nearby contaminants.  \\

\emph{EC 13140-1520:}
Also known as LP 737-47 and WD 1314-153. This object has a tenuous excess, and no high spatial resolution imaging exists to search for nearby contaminants.  \\

\emph{EC 20036-6613:}
The white dwarf model is well constrained by the multiple optical photometry data points but the object is lacking in near-infrared photometry. We obtained follow-up \emph{K$_{s}$} band imaging with the SPARTAN infrared camera on the SOAR telescope, which shows a nearby source likely contaminating the \emph{WISE} photometry. \\

\emph{EC 21010-1741:}
The white dwarf model is well constrained by the high quality VISTA infrared data points, but the infrared excess is not consistent with a single source between \emph{W1} and \emph{W2}. The VHS \emph{K$_{s}$} band image available shows a potential nearby contaminant, which is likely the source of the infrared excess. \\

\emph{EC 21459-3548:}
The white dwarf model is not well constrained by the limited available photometry, particularly in the near-infrared. ATLAS \emph{z} band imaging reveals a nearby contaminant that is too close to be resolved in \emph{WISE} photometry. \\

\emph{EC 23379-3725}
The best-fitted single temperature blackbody parameters place it well away from the overlapping stellar models. The discrepant near-infrared data present some concern for contamination, but the VHS \emph{K$_{s}$} band imaging is free of nearby contaminants. \\

\emph{Rejected Candidates:} We chose to reject 9 candidate dusty debris white dwarfs for a variety of reasons. Their SEDs are shown in Figure 10. EC 00323-3146 and EC 04552-2812 were designated spectral type 'DAwk' in the EC catalog, indicating narrow/weak Balmer lines preventing clear DA white dwarf classification. EC 19442-4207 is identified in the EC catalog as a CV/DAe, indicating the system is a cataclysmic variable \citep{odo13}. Both EC 12303-3052 and EC 11023-1821 were followed up with \emph{Spitzer} by \cite{roc15} and found to have no infrared excess, indicating the \emph{WISE} excesses are the result of contamination. EC 04114-1243 and EC05024-5705 both have white dwarf temperatures which are too high for optically thick dust to survive sublimation within their tidal disruption radius \citep{von07}. EC 04139-4029 and EC 04516-4428 are both are strong outliers in Figure 3. Their spectral energy distributions are suggestive of source confusion, particularly the highly discrepant near-infrared data from 2MASS and VHS. The higher spatial resolution images for both objects from VHS show nearby sources, indicating the \emph{WISE} excesses are very likely the result of contamination.

\newpage
\section{Example Spectral Energy Distributions with Single Temperature Blackbody Fits}
\begin{figure*}
\includegraphics[keepaspectratio=True,width=1.0\textwidth]{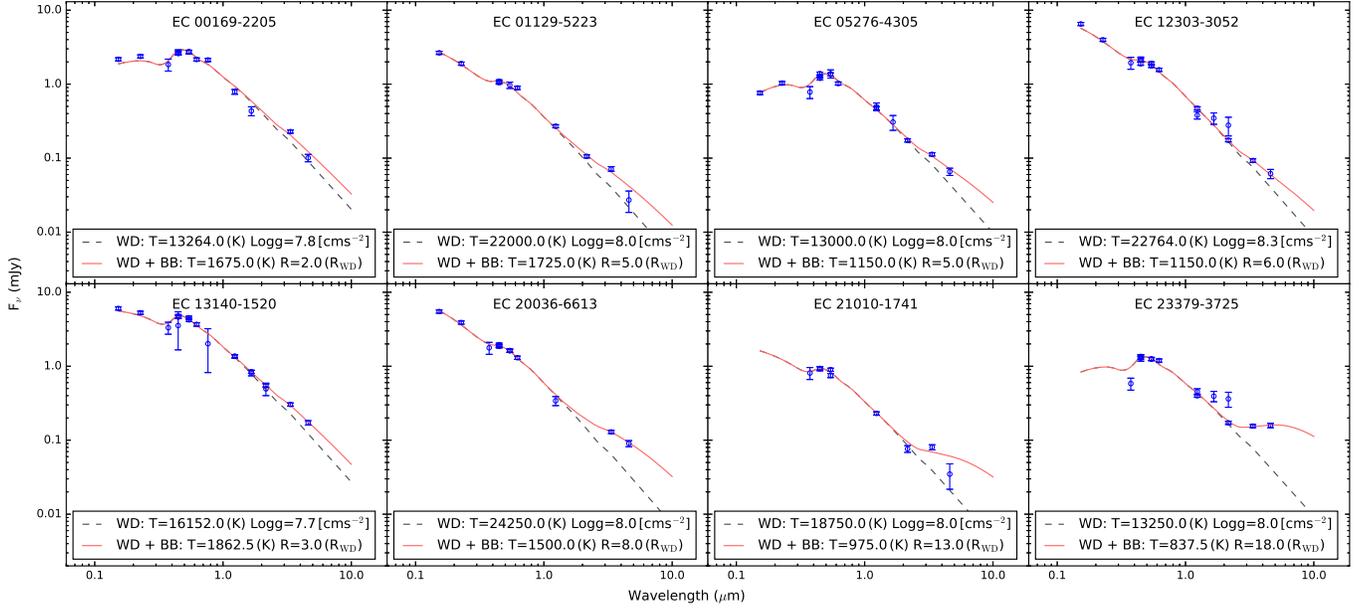}
\caption{SEDs of new dusty debris disk candidates not shown in Figure 5.}
\end{figure*}

\begin{figure*}
\includegraphics[keepaspectratio=True,width=1.0\textwidth]{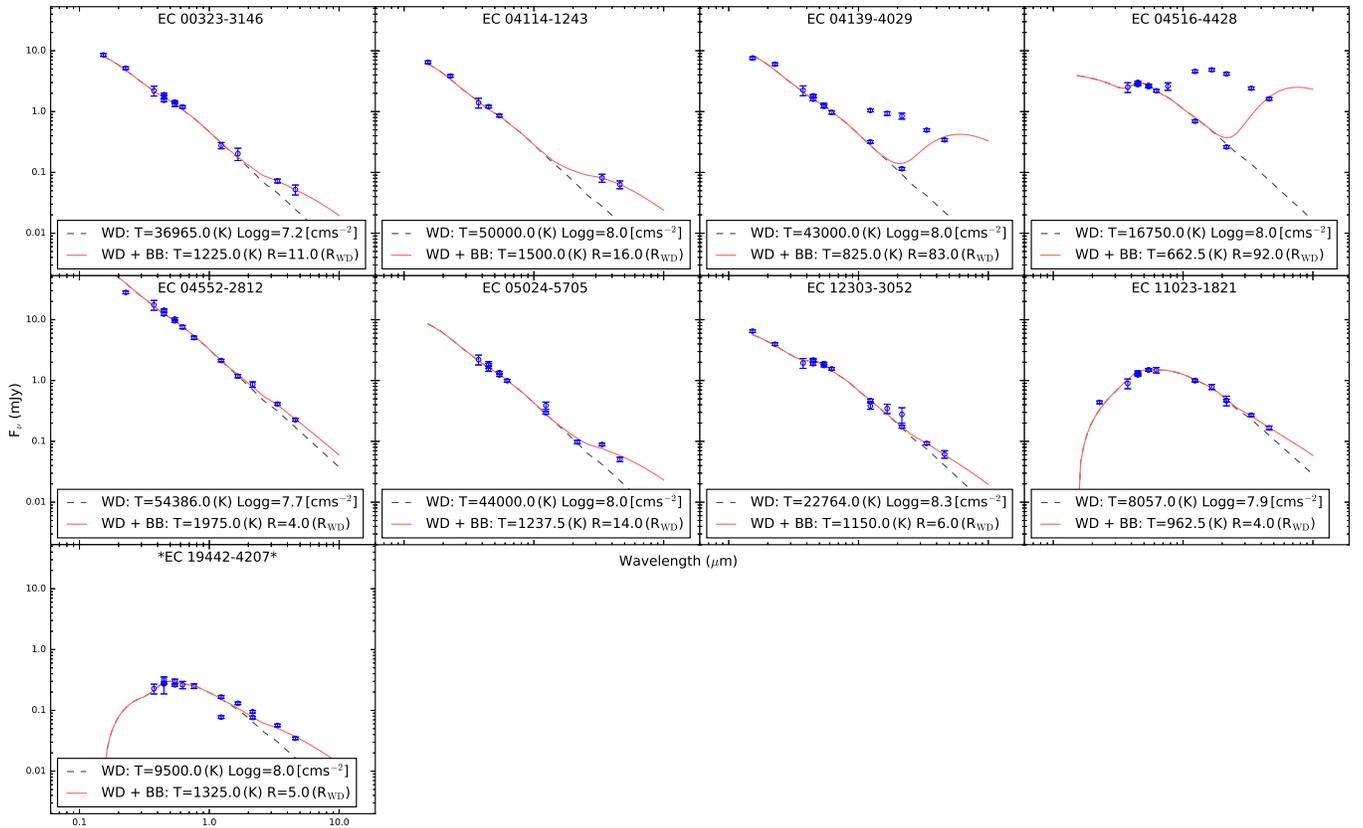}
\caption{SEDs of rejected candidates. *Denotes fluxes plotted in (Jy). The discrepant near-infrared photometry seen in EC 04139-4029, EC04516-4428, and EC19442-4207 results from source confusion in the 2MASS photometry that was resolved by the higher spatial resolution VHS photometry.}
\end{figure*}

\begin{figure*}
\includegraphics[keepaspectratio=True,width=1.0\textwidth]{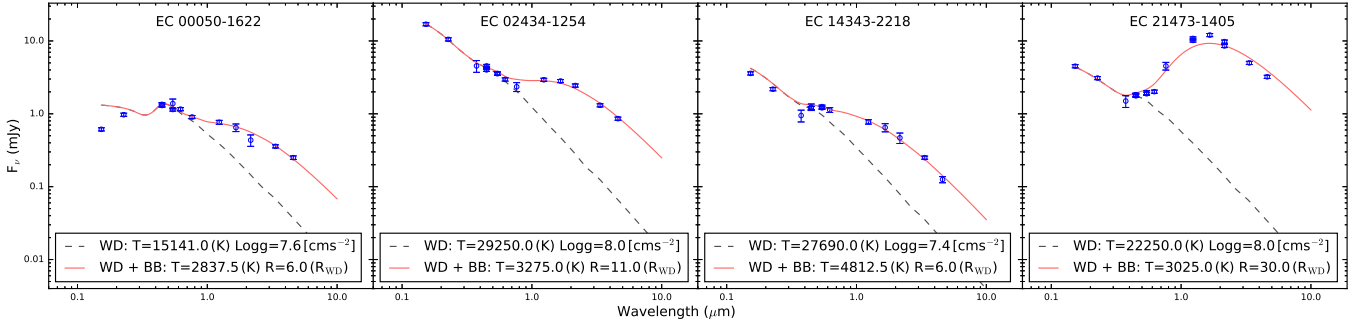}
\caption{Examples of SEDs of stellar/sub-stellar companions}
\end{figure*}

\begin{figure*}
\includegraphics[keepaspectratio=True,width=1.0\textwidth]{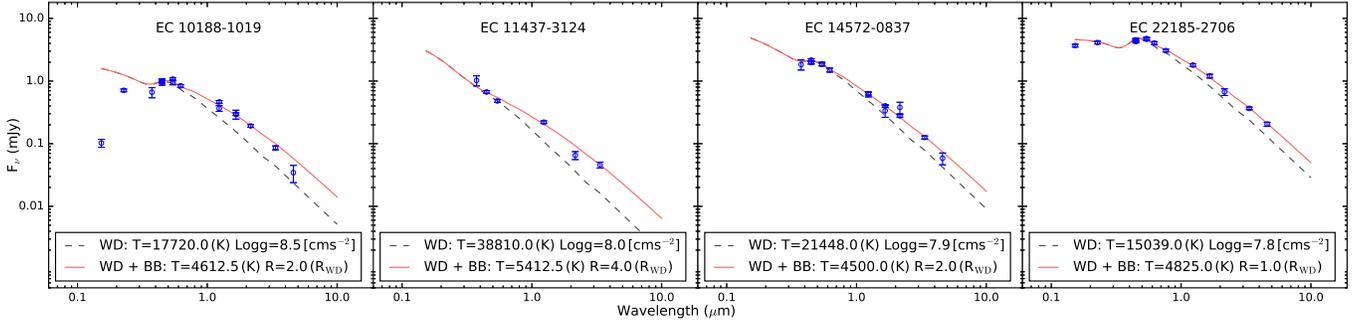}
\caption{Examples of SEDs with High Temperature/Low Radius Excesses}
\end{figure*}

\newpage
\section{Tables of Candidates from each Region}
\startlongtable
\begin{deluxetable*}{lcccccccccr}
\tablecaption{Region I: Dusty White Dwarf Candidates}
\tablecolumns{11}
\tablehead{\colhead{EC Name} & \colhead{Right Ascension} & \colhead{Declination} & \colhead{V} & \colhead{EC Sptype}& \colhead{WD T$_{\rm eff}$} & \colhead{WD $\log {g}$} & \colhead{BB T$_{\rm eff}$} & \colhead{BB Rad} & \colhead{Im Flag} &\colhead{Ref} \\ \colhead{} & \colhead{(J2000)} & \colhead{(J2000)} & \colhead{(mag)} & \colhead{} & \colhead{(K)} & \colhead{(cm$^{-2}$)} & \colhead{(K)} & \colhead{(R$_{\rm WD}$)} &\colhead{} &\colhead{}}
\startdata
\sidehead{New Candidates:}
00169-2205 & 4.8676032 & -21.817987 & 15.33 & DA & 13264.0\tablenotemark{a} & 7.78\tablenotemark{a} & 1675.0 & 2.0 & 20 & 1  \\ 
01071-1917 & 17.3880821 & -19.0216157 & 16.16 & DA & 14304.0\tablenotemark{a} & 7.79\tablenotemark{a} & 537.5 & 39.0 & 10 & 1  \\ 
01129-5223 & 18.7553121 & -52.1286677 & 16.47 & DA & 22000.0 & 8.0 & 1725.0 & 5.0 & 00 & 1  \\ 
02566-1802 & 44.7483175 & -17.8387548 & 16.51 & DA & 26212.0\tablenotemark{a} & 7.76\tablenotemark{a} & 575.0 & 49.0 & 10 & 1  \\ 
03103-6226 & 47.8357894 & -62.2545421 & 16.05 & DA & 16000.0 & 8.0 & 1050.0 & 11.0 & 00 & 1  \\ 
05276-4305 & 82.3005907 & -43.0595245 & 16.10 & DA & 13000.0 & 8.0 & 1150.0 & 5.0 & 00 & 1  \\ 
13140-1520 & 199.1818104 & -15.5976591 & 14.86 & DA3 & 16152.0\tablenotemark{a} & 7.72\tablenotemark{a} & 1862.5 & 3.0 & 20 & 1  \\ 
20036-6613 & 302.099844 & -66.0769588 & 15.91 & DA & 24250.0 & 8.0 & 1500.0 & 8.0 & 20 & 1  \\ 
21010-1741 & 315.9666489 & -17.4904601 & 16.57 & DA & 18750.0 & 8.0 & 975.0 & 13.0 & 01 & 1  \\ 
21548-5908 & 329.5997382 & -58.8983987 & 15.75 & DA & 12000.0 & 8.0 & 1037.5 & 9.0 & 00 & 1  \\ 
23379-3725 & 355.1531593 & -37.1454489 & 16.18 & DA & 13250.0 & 8.0 & 837.5 & 18.0 & 00 & 1  \\ 
\sidehead{Previously Identified:}
04203-7310 & 64.9071296 & -73.0622893 & 15.61 & DA & 19000.0 & 8.0 & 1125.0 & 19.0 & 20 & 2  \\ 
05365-4759 & 84.473051 & -47.9679045 & 15.63 & DA & 22250.0 & 8.0 & 1037.5 & 16.0 & 01 & 3 \\ 
11507-1519 & 178.3133581 & -15.6099161 & 16.00 & DA5 & 12132.0\tablenotemark{a} & 8.03\tablenotemark{a} & 862.5 & 31.0 & 20 & 4  \\ 
21159-5602 & 319.9006626 & -55.8370306 & 14.27 & DA & 9625.0\tablenotemark{a} & 8.01\tablenotemark{a} & 900.0 & 10.0 & 00 & 5  \\ 
22215-1631 & 336.0727067 & -16.263386 & 15.45 & DA & 9937.0\tablenotemark{a} & 8.16\tablenotemark{a} & 925.0 & 6.0 & 00 & 6  \\ 
\sidehead{Rejected Candidates:}
00323-3146 & 8.7072847 & -31.4978588 & 16.09 & DAwk & 36965.0\tablenotemark{a} & 7.19\tablenotemark{a} & 1225.0 & 11.0 & 10 & 1  \\ 
04114-1243 & 63.4385557 & -12.5944927 & 16.70 & DA & 50000.0 & 8.0 & 1500.0 & 16.0 & 20 & 1  \\ 
04139-4029 & 63.9157837 & -40.3757732 & 16.23 & DA & 43000.0 & 8.0 & 825.0 & 83.0 & 01 & 1  \\ 
04516-4428 & 73.3031017 & -44.394365 & 15.35 & DA & 16750.0 & 8.0 & 662.5 & 92.0 & 01 & 1  \\ 
04552-2812 & 74.3051311 & -28.1312704 & 13.98 & DAwk & 54386.0\tablenotemark{a} & 7.68\tablenotemark{a} & 1975.0 & 4.0 & 20 & 7  \\ 
05024-5705 & 75.8461981 & -57.0227396 & 16.22 & DA & 44000.0 & 8.0 & 1237.5 & 14.0 & 01 & 1  \\ 
11023-1821 & 166.1945881 & -18.6200168 & 15.99 & DA5 & 8057.0\tablenotemark{a} & 7.85\tablenotemark{a} & 962.5 & 4.0 & 20 & 8  \\ 
12303-3052 & 188.2520412 & -31.1432902 & 15.81 & DA2 & 22764.0\tablenotemark{a} & 8.28\tablenotemark{a} & 1150.0 & 6.0 & 20 & 1  \\ 
19442-4207 & 296.9186198 & -42.0074873 & 10.38 & CV/DAe & 9500.0 & 8.0 & 1325.0 & 5.0 & 01 & 9  \\ 
\enddata
\tablenotetext{a}{Spectroscopic parameters from \cite{koe09}}
\tablenotetext{b}{Spectroscopic parameters from \cite{gia11}}
\tablecomments{References: (1) This Paper; (2) \cite{hoa13}; (3) \cite{den16}; (4) \cite{kil07}; (5) \cite{von07}; (6) \cite{far10}; (7) \cite{dob05}; (8) \cite{far08}; (9) \cite{odo13}}
\tablecomments{Under the column Im Flag, the first bit refers to the quality of imaging, with objects that have VHS \emph{K$_{s}$} band-images receiving a 0, VST-ATLAS \emph{z} band images a 1, and those without follow-up images a 2. The second bit refers to the potential for contamination. Objects which appeared as single stars were assigned a 0, those with one or more potential contaminants within the 7.8\arcsec\, circle were assigned a 1.}
\end{deluxetable*}

\startlongtable
\begin{deluxetable*}{lcccccccccr}
\tablecaption{Region II: White Dwarfs with Stellar/Sub-Stellar Excesses}
\tablecolumns{11}
\tablehead{\colhead{EC Name} & \colhead{Right Ascension} & \colhead{Declination} & \colhead{V} & \colhead{EC Sptype}& \colhead{WD T$_{\rm eff}$} & \colhead{WD $\log {g}$} & \colhead{BB T$_{\rm eff}$} & \colhead{BB Rad} & \colhead{Im Flag} &\colhead{Ref} \\ \colhead{} & \colhead{(J2000)} & \colhead{(J2000)} & \colhead{(mag)} & \colhead{} & \colhead{(K)} & \colhead{(cms$^{-2}$)} & \colhead{(K)} & \colhead{(R$_{\rm WD}$)} &\colhead{} &\colhead{}}
\startdata
\sidehead{New Candidates:}
00050-1622 & 1.8951187 & -16.0922234 & 16.29 & DA & 15141.0\tablenotemark{a} & 7.59\tablenotemark{a} & 2837.5 & 6.0 & 10 & 1  \\ 
00166-4340 & 4.775084 & -43.4051481 & 15.53 & DA & 7250.0\tablenotemark{a} & 8.0\tablenotemark{a} & 2812.5 & 6.0 & 00 & 1  \\ 
00286-6338 & 7.7279283 & -63.3624649 & 15.23 & DA & 19750.0 & 8.0 & 3225.0 & 10.0 & 00 & 1  \\ 
00370-4201 & 9.8542309 & -41.7470551 & 16.37 & DA & 10750.0 & 8.0 & 3875.0 & 1.0 & 00 & 1  \\ 
00594-5701 & 15.3797183 & -56.7644763 & 16.55 & DA & 13000.0 & 8.0 & 3187.5 & 9.0 & 00 & 1  \\ 
01077-8047 & 17.0733956 & -80.5236636 & 14.47 & DA & 5500.0 & 8.0 & 2912.5 & 4.0 & 20 & 1  \\ 
01176-8233 & 19.3494814 & -82.3011543 & 16.43 & DA & 17750.0 & 8.0 & 3737.5 & 17.0 & 20 & 1  \\ 
01346-4042 & 24.2001527 & -40.4593443 & 16.36 & DA/DAB & 13500.0 & 8.0 & 2950.0 & 17.0 & 00 & 1  \\ 
02223-2630 & 36.1504952 & -26.2812935 & 15.68 & DA & 23198.0\tablenotemark{a} & 7.91\tablenotemark{a} & 2087.5 & 6.0 & 10 & 2  \\ 
02434-1254 & 41.4726889 & -12.7056873 & 15.05 & DA & 29250.0 & 8.0 & 3275.0 & 11.0 & 10 & 1  \\ 
03155-1747 & 49.4484126 & -17.601521 & 16.48 & DA & 22750.0 & 8.0 & 2875.0 & 32.0 & 10 & 1  \\ 
03378-8348 & 53.0676708 & -83.6389586 & 16.24 & DA & 36750.0 & 8.0 & 3387.5 & 11.0 & 20 & 1  \\ 
04094-3233 & 62.838107 & -32.4373756 & 16.01 & DA & 18250.0\tablenotemark{a} & 8.0\tablenotemark{a} & 3037.5 & 16.0 & 20 & 1  \\ 
04233-2822 & 66.3363893 & -28.255434 & 16.48 & DA & 10907.0\tablenotemark{a} & 8.07\tablenotemark{a} & 2087.5 & 6.0 & 20 & 1  \\ 
04310-3259 & 68.2266693 & -32.8872894 & 17.6 & DA & 3750.0 & 8.0 & 2600.0 & 2.0 & 20 & 1  \\ 
04365-1633 & 69.6966934 & -16.4545871 & 16.03 & DA & 14092.0\tablenotemark{a} & 7.96\tablenotemark{a} & 2825.0 & 2.0 & 20 & 1  \\ 
04567-2347 & 74.714622 & -23.7150737 & 16.62 & DA & 23645.0\tablenotemark{a} & 7.79\tablenotemark{a} & 3175.0 & 5.0 & 20 & 1  \\ 
05089-5933 & 77.4280658 & -59.4939338 & 15.78 & DA & 28500.0 & 8.0 & 3750.0 & 61.0 & 00 & 1  \\ 
05230-3821 & 81.1923171 & -38.3099344 & 16.55 & DA & 18250.0 & 8.0 & 2712.5 & 11.0 & 20 & 1  \\ 
05237-3856 & 81.3667125 & -38.903283 & 16.17 & DA & 15750.0\tablenotemark{a} & 8.0\tablenotemark{a} & 2937.5 & 15.0 & 20 & 1  \\ 
05387-3558 & 85.1301338 & -35.9572189 & 13.97 & DA & 13250.0 & 8.0 & 3925.0 & 2.0 & 20 & 1  \\ 
05430-4711 & 86.09625 & -47.1715794 & 15.97 & DA & 8000.0 & 8.0 & 2650.0 & 9.0 & 00 & 1  \\ 
12204-2915 & 185.7709471 & -29.5410766 & 15.79 & DA3 & 17702.0\tablenotemark{a} & 7.89\tablenotemark{a} & 2387.5 & 2.0 & 20 & 1  \\ 
13123-2523 & 198.7660094 & -25.6497229 & 15.69 & DA1 & 75463.0\tablenotemark{a} & 7.68\tablenotemark{a} & 3425.0 & 17.0 & 20 & 1  \\ 
13324-2255 & 203.7936553 & -23.1771076 & 16.30 & DA3 & 20264.0\tablenotemark{a} & 7.86\tablenotemark{a} & 2125.0 & 6.0 & 20 & 1  \\ 
14265-2737 & 217.3638143 & -27.8498806 & 15.92 & DA3 & 18087.0\tablenotemark{a} & 7.66\tablenotemark{a} & 3062.5 & 2.0 & 20 & 1  \\ 
14361-1832 & 219.744523 & -18.7615606 & 16.56 & DA? & 29250.0 & 8.0 & 3737.5 & 8.0 & 20 & 1  \\ 
19272-7152 & 293.2369964 & -71.7669411 & 15.92 & DA & 20250.0 & 8.0 & 3525.0 & 12.0 & 20 & 1  \\ 
20453-7549 & 312.7909386 & -75.6400976 & 16.05 & DA & 25750.0 & 8.0 & 2500.0 & 3.0 & 20 & 1  \\ 
20503-4650 & 313.4347278 & -46.6575692 & 15.76 & DAwk & 19250.0 & 8.0 & 4100.0 & 10.0 & 01 & 1  \\ 
21053-8201 & 318.3144577 & -81.8191237 & 13.63 & DA & 10600.0\tablenotemark{a} & 8.24\tablenotemark{a} & 3587.5 & 1.0 & 20 & 1  \\ 
21105-5128 & 318.4991085 & -51.2753573 & 16.68 & DA & 16500.0 & 8.0 & 3262.5 & 4.0 & 01 & 1  \\ 
21161-2610 & 319.7694033 & -25.9702383 & 16.10 & DA & 24750.0 & 8.0 & 3400.0 & 2.0 & 01 & 1  \\ 
21188-2715 & 320.4373582 & -27.0364774 & 15.16 & DA & 5250.0 & 8.0 & 2625.0 & 6.0 & 20 & 1  \\ 
21335-3637 & 324.162532 & -36.4000738 & 15.73 & DA & 26940.0\tablenotemark{b} & 7.75\tablenotemark{b} & 3725.0 & 2.0 & 10 & 1  \\ 
21459-3548 & 327.2267922 & -35.5801944 & 16.35 & DA & 12000.0 & 8.0 & 2400.0 & 3.0 & 11 & 1  \\ 
21470-5412 & 327.6000537 & -53.9776573 & 15.26 & DA & 11500.0 & 8.0 & 2875.0 & 23.0 & 00 & 1  \\ 
21473-1405 & 327.5153847 & -13.8626911 & 15.75 & DA & 22250.0\tablenotemark{a} & 8.0\tablenotemark{a} & 3025.0 & 30.0 & 00 & 1  \\ 
22016-3015 & 331.1442181 & -30.0183606 & 15.58 & DAe & 12750.0 & 8.0 & 2762.5 & 31.0 & 10 & 1  \\ 
22158-2027 & 334.6492133 & -20.2113139 & 16.00 & DA & 15500.0 & 8.0 & 3750.0 & 2.0 & 20 & 1  \\ 
23016-4857 & 346.12937 & -48.6825705 & 15.58 & DA & 4750.0 & 8.0 & 2862.5 & 2.0 & 00 & 1  \\ 
23227-6739 & 351.4332031 & -67.3785049 & 16.71 & DAwk & 31500.0 & 8.0 & 2900.0 & 9.0 & 20 & 1  \\ 
\sidehead{Previously Identified:}
00370-6328 & 9.8125271 & -63.2073005 & 15.82 & *DAe sdB+G & 50000.0 & 8.0 & 3175.0 & 41.0 & 00 & 3,4  \\ 
01162-2310 & 19.6547306 & -22.9156167 & 16.15 & DA & 31990.0\tablenotemark{b} & 7.63\tablenotemark{b} & 3225.0 & 20.0 & 11 & 5  \\ 
01319-1622 & 23.6002247 & -16.1189799 & 13.94 & DA & 50110.0\tablenotemark{a} & 7.87\tablenotemark{a} & 3337.5 & 18.0 & 10 & 5  \\ 
01450-2211 & 26.8410187 & -21.9475691 & 14.85 & DA(Z) & 11747.0\tablenotemark{a} & 8.07\tablenotemark{a} & 2387.5 & 5.0 & 10 & 5  \\ 
01450-7035 & 26.5471352 & -70.339152 & 15.77 & DA & 19000.0 & 8.0 & 3375.0 & 6.0 & 20 & 5  \\ 
02083-1520 & 32.6787961 & -15.1095691 & 15.19 & DA+dM & 22620.0\tablenotemark{b} & 7.92\tablenotemark{b} & 3250.0 & 28.0 & 11 & 5  \\ 
03094-2730 & 47.888501 & -27.3236914 & 15.70 & DA ? & 56610.0\tablenotemark{b} & 7.53\tablenotemark{b} & 3712.5 & 40.0 & 10 & 5,6  \\ 
03319-3541 & 53.4679871 & -35.5210623 & 14.38 & DA+dMe & 23000.0\tablenotemark{a} & 8.0\tablenotemark{a} & 3200.0 & 31.0 & 10 & 4,5  \\ 
03338-6410 & 53.643096 & -64.0156465 & 14.32 & DA & 50000.0 & 8.0 & 3050.0 & 34.0 & 20 & 7  \\ 
03479-1344 & 57.5607475 & -13.5872177 & 14.96 & DA & 14250.0\tablenotemark{a} & 7.76\tablenotemark{a} & 2862.5 & 23.0 & 11 & 5  \\ 
03569-2320 & 59.7703036 & -23.2070023 & 15.87 & DAwk & 74710.0\tablenotemark{b} & 7.86\tablenotemark{b} & 3625.0 & 17.0 & 11 & 5  \\ 
10150-1722 & 154.3701471 & -17.6187202 & 16.77 & DA4 & 32370.0\tablenotemark{b} & 7.58\tablenotemark{b} & 4075.0 & 15.0 & 20 & 5,6  \\ 
12477-1738 & 192.5921204 & -17.9129074 & 16.20 & DA+dMe & 21620.0\tablenotemark{a} & 8.16\tablenotemark{a} & 3287.5 & 27.0 & 20 & 5,8  \\ 
12540-1318 & 194.1649138 & -13.5783955 & 16.04 & DA2 & 23710.0\tablenotemark{b} & 7.92\tablenotemark{b} & 3325.0 & 14.0 & 20 & 5  \\ 
13077-1411 & 197.5938438 & -14.4525458 & 16.44 & DA4 & 26400.0\tablenotemark{b} & 7.92\tablenotemark{b} & 3737.5 & 25.0 & 20 & 5,6  \\ 
13198-2849 & 200.6679344 & -29.0922276 & 15.99 & DA+dM & 16620.0\tablenotemark{a} & 7.75\tablenotemark{a} & 2925.0 & 34.0 & 20 & 5,8  \\ 
13349-3237 & 204.4613066 & -32.872784 & 16.34 & hot DA? sd? & 7000.0\tablenotemark{a} & 8.0\tablenotemark{a} & 3262.5 & 4.0 & 20 & 9  \\ 
13471-1258 & 207.4669075 & -13.2272276 & 14.80 & DA+dM & 3000.0\tablenotemark{a} & 8.0\tablenotemark{a} & 2612.5 & 2.0 & 20 & 8  \\ 
14329-1625 & 218.9405799 & -16.63817 & 14.89 & DA+dMe & 16500.0\tablenotemark{a} & 8.0\tablenotemark{a} & 2837.5 & 27.0 & 20 & 8,9  \\ 
14363-2137 & 219.8024563 & -21.8369585 & 15.94 & DA6 & 23690.0\tablenotemark{a} & 7.91\tablenotemark{a} & 3225.0 & 31.0 & 20 & 5  \\ 
20220-2243 & 306.2477455 & -22.5559774 & 16.45 & DAwk/cont & 50000.0 & 8.0 & 2200.0 & 12.0 & 20 & 10  \\ 
20246-4855 & 307.0651962 & -48.7608551 & 15.69 & DA+dM & 16250.0 & 8.0 & 3175.0 & 31.0 & 00 & 11  \\ 
21016-3627 & 316.1957085 & -36.2570282 & 16.79 & DA & 50000.0 & 8.0 & 4137.5 & 26.0 & 20 & 5  \\ 
21083-4310 & 317.9062002 & -42.9697055 & 15.77 & DA & 7750.0 & 8.0 & 2787.5 & 10.0 & 00 & 5  \\ 
21384-6423 & 325.593066 & -64.1624192 & 15.87 & DA+dMe & 17750.0 & 8.0 & 2962.5 & 30.0 & 01 & 3,4  \\ 
22049-5839 & 332.0912832 & -58.4093375 & 14.22 & DA+dM & 19500.0 & 8.0 & 3475.0 & 26.0 & 00 & 3,4  \\ 
23260-2226 & 352.1615109 & -22.1721571 & 16.80 & DA & 19590.0\tablenotemark{b} & 8.01\tablenotemark{b} & 3212.5 & 32.0 & 11 & 5  \\ 
\enddata
\tablenotetext{a}{Spectroscopic parameters from \cite{koe09}}
\tablenotetext{b}{Spectroscopic parameters from \cite{gia11}}
\tablecomments{References: (1) This Paper; (2) \cite{roc15} (3) \cite{kil15}; (4) \cite{kil16}; (5) \cite{hoa07}; (6) \cite{far10}; (7) \cite{tap07}; (8) \cite{kil97}; (9) \cite{tap07}; (10) \cite{dow01}; (11) \cite{odo13}}
\end{deluxetable*}

\startlongtable
\begin{deluxetable*}{lcccccccccr}
\tablecaption{Region III: White Dwarfs with High Temperature/Low Radius Excesses}
\tablecolumns{11}
\tablehead{\colhead{EC Name} & \colhead{Right Ascension} & \colhead{Declination} & \colhead{V} & \colhead{EC Sptype}& \colhead{WD T$_{\rm eff}$} & \colhead{WD $\log {g}$} & \colhead{BB T$_{\rm eff}$} & \colhead{BB Rad} & \colhead{Im Flag} &\colhead{Ref} \\ \colhead{} & \colhead{(J2000)} & \colhead{(J2000)} & \colhead{(mag)} & \colhead{} & \colhead{(K)} & \colhead{(cm$^{-2}$)} & \colhead{(K)} &  \colhead{(R$_{\rm WD}$)} &\colhead{} &\colhead{}}
\startdata
00169-3216 & 4.8518404 & -31.9982455 & 15.67 & sdB/DA? & 30750.0 & 8.0 & 4287.5 & 5.0 & 10 & 1  \\ 
02121-5743 & 33.4391834 & -57.4967862 & 14.34 & DAwk & 17000.0 & 8.0 & 4625.0 & 2.0 & 00 & 1  \\ 
03120-6650 & 48.1736862 & -66.6561967 & 16.73 & DA/sdB & 24250.0 & 8.0 & 5125.0 & 2.0 & 20 & 1  \\ 
03372-5808 & 54.5997256 & -57.9739414 & 16.45 & DA & 44250.0 & 8.0 & 5500.0 & 3.0 & 00 & 1  \\ 
03572-5455 & 59.6228876 & -54.7779686 & 16.09 & sdB?/DA? & 22250.0 & 8.0 & 5287.5 & 3.0 & 00 & 1  \\ 
04536-2933 & 73.8992797 & -29.4835602 & 14.97 & DAB & 20640.0\tablenotemark{b} & 7.61\tablenotemark{b} & 4925.0 & 1.0 & 20 & 1  \\ 
10188-1019 & 155.3301617 & -10.5804209 & 16.35 & DA5 & 17720.0\tablenotemark{b} & 8.52\tablenotemark{b} & 4612.5 & 2.0 & 20 & 1  \\ 
11437-3124 & 176.575735 & -31.6839625 & 17.32 & DA1 & 38810.0\tablenotemark{b} & 8.04\tablenotemark{b} & 5412.5 & 4.0 & 20 & 1  \\ 
14572-0837 & 224.9707361 & -8.8247843 & 15.77 & DA2 & 21448.0\tablenotemark{a} & 7.92\tablenotemark{a} & 4500.0 & 2.0 & 20 & 2  \\ 
19579-7344 & 300.9522623 & -73.5956704 & 16.65 & DA & 25250.0 & 8.0 & 4975.0 & 3.0 & 20 & 1  \\ 
20228-5030 & 306.6264874 & -50.3451028 & 17.13 & DA & 21250.0 & 8.0 & 5500.0 & 3.0 & 20 & 1  \\ 
21591-7353 & 330.8978119 & -73.6455014 & 14.46 & DA & 19750.0 & 8.0 & 4812.5 & 3.0 & 20 & 1  \\ 
22185-2706 & 335.3493648 & -26.8484764 & 14.76 & DA & 15039.0\tablenotemark{a} & 7.8\tablenotemark{a} & 4825.0 & 1.0 & 10 & 1  \\ 
23127-4239 & 348.8769107 & -42.392647 & 16.24 & sdB/DAwk & 29750.0 & 8.0 & 5500.0 & 2.0 & 00 & 1  \\ 
23513-5536 & 358.4795859 & -55.3316608 & 16.23 & sdB/DA & 24000.0 & 8.0 & 5275.0 & 3.0 & 00 & 1  \\ 
\enddata
\tablenotetext{a}{Spectroscopic parameters from \cite{koe09}}
\tablenotetext{b}{Spectroscopic parameters from \cite{gia11}}
\tablecomments{References: (1) This Paper; (2) \cite{far09}}
\end{deluxetable*}

\end{document}